\documentclass[final]{aipproc}
\layoutstyle{6x9}
\usepackage{graphicx, color}
\usepackage{amsmath}

\usepackage[english]{babel}
\usepackage[latin1]{inputenc}

\begin{document}

\title{Ultra High Energy Cosmic Ray, Neutrino, and Photon Propagation and the Multi-Messenger Approach}


\author{Andrew Taylor}{
  address={Max-Planck-Institut für Kernphysik\\
Postfach 10 39 80, D-69029 Heidelberg, Germany}}

\author{Alexandra De Castro}{
  address={Departamento de Física, Universidad Simón Bolívar\\
Ap 89000, Caracas 1080-A, Venezuela}}

\author{Edith Castillo-Ruiz}{
 address={Secci\'on F\'{\i}sica, Departamento de Ciencias, Pontificia
Universidad Cat\'{o}lica del Per\'{u},\\ 
Apartado 1761, Lima, Per\'{u}}} 

\begin{abstract}
The propagation of UHECR nuclei for $A=1$ (protons) to $A=56$ (iron) from cosmological sources 
through extragalactic space is discussed in the first lecture.
This is followed in the second and third lectures by a consideration of the
generation and propagation of secondary particles produced via the UHECR loss interactions.
In the second lecture we focus on the generation of the diffuse cosmogenic UHE-neutrino flux.
In the third lecture we investigate the arriving flux of UHE-photon flux at Earth.
In the final lecture the results of the previous lectures are put together in order to provide
new insights into UHECR sources. 
The first of these providing a means with which to investigate the local population of UHECR sources
through the measurement of the UHECR spectrum and their photon fraction at Earth.
The second of these providing contraints on the UHECR source radiation fields through the possible
observation at Earth of UHECR nuclei.
\end{abstract}

\maketitle

\section{Lecture 1: UHECR Proton and Nuclei Propagation}
Over the last century, Cosmic Rays (CR) have been detected arriving at Earth. The energy range with which
these particles arrive spans over ten decades in energy from $10^{9}$~eV to $10^{20}$~eV. 
For a large part of this energy range, the CR spectrum is found to fit a polynomial description 
of the form $dN/dE\propto E^{-\alpha}$, with $\alpha$ the spectral index, taking value of roughly $2.7$ up to 
$10^{15}$~eV (1~particle~m$^{2}$~yr$^{-1}$), changing above this energy to $\alpha=3.1$ 
(producing a feature called the ``knee'' \cite{Antoni:2005wq}). Further on, at higher energies, 
a second ``knee'' in the spectrum 
is observed near $10^{17}$~eV. Following this, at energies just above $10^{18}$~eV (1~particle~km$^{2}$~yr$^{-1}$),
another sudden change occurs in the spectrum, with it apparently returning to a spectral index value of $\alpha=2.7$ 
(the ``ankle''). CR with energies over $10^{18}$~eV are referred to as
{\it Ultra High Energy Cosmic Rays} (UHECR). From a simple consideration of the Larmor radius of CR
in the Galaxy it is seen that UHECR containment within the Galaxy's $\mu$G magnetic fields is problematic, 
most probably requiring their sources to be extragalactic in origin at these energies. 
Being charged particles, UHECR will execute trajectories with non-negligible curvature in nG extragalactic 
magnetic fields present on distance scales $>$Mpc from their sources \cite{Hooper:2006tn}.
This would lead to a washing out of their directional (source) information on such scales. 

With direct CR detection only being possible in 
the low energy region $<10^{14}$~eV, where the arriving fluxes are sufficiently large to allow satellite and balloon 
spectroscopy, a direct handle on the CR composition is only possible for these lower energy particles. 
At higher energies only indirect measurements are viable through the analysis of the 
profile and content of the particle shower created in the atmosphere. However, for the hadronic part of the 
shower, the composition of the primary particle plays an important role in the shower development, dictating 
the eventual shower profile and content. Thus, the composition
information is encoded in the CR showers detected.

In this first lecture we consider the propagation of UHECR nuclei for $A=1$ (protons) to $A=56$ (iron)
from cosmological sources through interstellar space to Earth. Firstly, the physics 
governing the propagation of UHE-protons will be discussed. This is followed in the second section 
by a discussion of nuclei UHECR propagation. In the third section we study the 
UHECR spectra observed at Earth for different primary particles injected at the sources. Finally, we
end the lecture with a brief description of an analytic approach for UHECR nuclei propagation.

\section{1.2 Composition of Cosmic Rays}
As already mentioned, the flux of UHECR is sufficiently low that a present-day satellite 
born flux measurement, if attempted, would need to wait longer than $\sim 10^{6}$~yrs just to detect one.
Instead, in order to detect a significant amount of events it is necessary to use the Earth's atmosphere
as the detector, as employed by current day UHECR air shower detectors such as the Pierre Auger Observatory
\cite{Abraham:2004dt} (Auger) in Argentina. In these experiments, indirect detection of UHECR is achieved via the secondary 
particles generated in a particle shower created through the interaction of a (primary) CR
with the nuclei of atoms in the atmosphere. The price payed through this approach, with present uncertainty 
in the description of the hadronic shower development, is the loss of precise composition information of the 
primary particle. Nevertheless, there are some particle shower 
characteristics which do intimately relate to the primary particle composition, they are: 
\begin{itemize}
\item the distance from the "top of the atmosphere", measured in column depth, at which the number of particles 
in the shower is maximum ($X_{\rm max}$), $X_{\rm max}^{p}(E)>X_{\rm max}^{Fe}(E)$
\item the number of muons in the shower ($N_{\mu}$), $N_{\mu}^{Fe}(E)>N_{\mu}^{p}(E)$
\end{itemize}

We here focus on the first of these composition indicators. 
The energy of the shower particles energy falls below a critical value $E_{\rm crit.}$. Through a Heitler 
model \cite{Heitler} description for the electromagnetic shower, with primary energy $E_{0}$,
the conservation of energy in the shower requires,
$E_{0}=E_{\rm crit.}2^{n_{\rm crit.}}$, where $n_{\rm crit.}$ is the number of particles in the shower
when the particles in the shower have energy $E_{\rm crit.}$. This  
$X_{\rm max}=n_{\rm crit.}\lambda\ln(2)=\lambda \ln(E_{0}/E_{\rm crit.})$ 
where $\lambda$ is the mean free path.
Typically $E_{\rm crit.}=85$~MeV for electromagnatic showers ($\gamma/e$).

Fig.~3 in ref.~\cite{Unger:2007mc} demonstates that recent Auger $X_{\rm max}$ measurements presently
support arguments for a mixed composition of UHECR arriving at Earth.
With such measurements as motivation, along with UHECR proton propagation we also review 
the main aspects of UHECR nuclei propagation in the following sections. Indeed, it is worth stating that the CR composition 
observed at Earth may be quite different from that at injection. Intermediate mass or heavy nuclei injected in 
a distant CR acceleration will gradually disintegrate into lighter nuclei and nucleons as they propagate 
through intergalactic space \cite{Hooper:2006tn}. 

\section{1.3 UHECR Proton Energy Loss Interactions}
The photon targets for pion production are given by the Cosmic microwave background (CMB)
and the Cosmic infrared background (CIR).
The specific number density of the CMB photon spectrum constitutes the main target for $p\gamma$ interactions
and is described by the Planck's law for a black body of temperature $kT=2.3\times 10^{-4}$~eV (2.7~K). 

\begin{equation}
\epsilon_{\gamma}\frac{dn_\gamma}{d\epsilon_\gamma}  = \frac{8\pi}{h^3c^3}\frac{ \epsilon^3_\gamma}{e^{\epsilon_\gamma/kT}-1}
\approx 170 \left[ \frac{\left(\epsilon_{\gamma}/kT\right)^3}{e^{\epsilon_{\gamma}/kT}-1} \right] ~{\rm cm}^{-3},\label{Espectro_BB}
\end{equation}

\noindent where $c$ is the velocity of light in a vacuum and $h$ is Planck's constant.

Over large enough distance scales through this background, UHECR protons will undergo interactions with these photons 
leading to the production of electron positron pairs ($p+\gamma \rightarrow p+e^{+}e^{-}$) and pions ($p+\gamma \rightarrow p/n+\pi^{0}/\pi^{+}$).

The attenuation rate calculation for both these ($p\gamma$) proton energy loss processes may be expressed as
\begin{equation}
R=\frac{1}{2\Gamma_{p}^{2}}\int_{0}^{\infty}\frac{1}{\epsilon_{\gamma}^{2}}\frac{dn_{\gamma}}{d\epsilon_{\gamma}}d\epsilon_{\gamma}\int_{0}^{2\Gamma_{p}\epsilon_{\gamma}}\epsilon_{\gamma}^{\prime} \sigma_{p\gamma}(\epsilon_{\gamma}^{\prime}) K_{p}d\epsilon_{\gamma}^{\prime}
\end{equation}
where $\Gamma=E_{p}/m_{p}$, $m_{p}$ is the protons mass, $\epsilon_{\gamma}$ is the photon energy, $\sigma_{p\gamma}$ is the 
$p\gamma$ interaction cross section and $K_p$ is the inelasticity of the proton for such interactions
($K_{p}=\Delta E_{p}/E_{p}$).

\noindent {\bf Pair production}- In the rest frame of the proton, the photon threshold energy for 
pair creation is $\epsilon'_{\gamma,th} \approx 1$~MeV. The inelasticities of these interactions typically going as 
$K_{p}\approx 4m_{e}^{2}c^{4}/(m_{p}c^{2}\epsilon_{\gamma}^{\prime})$, where $\epsilon_{\gamma}^{\prime}$ is the 
colliding photon's energy in the proton's rest frame and $m_{e}$ is the electron rest mass \cite{Chodorowski:1992}.

\noindent {\bf Pion production}-In the rest frame of the proton, the photon threshold energy for 
pion production is $\epsilon'_{\gamma,th} \approx 145$~MeV. The inelasticities of these interactions, for low pion multiplicities,
go approximately as $K_{p}\approx (m_{\pi}^{2}+2m_{p}\epsilon_{\gamma}^{\prime})/2(m_{p}^{2}+2m_{p}\epsilon_{\gamma}^{\prime})$, where $\epsilon_{\gamma}^{\prime}$ is
the colliding photon's energy in the proton rest frame, and $m_{\pi}$ is the pion rest mass \cite{Stecker:1968}.

For an approximate description of $R(E_{p})$ for pion losses, we may describe the cross-section as a 
top-hat function, 

\begin{equation} \label{ApproxAttnRate}
  \sigma(\epsilon'_{\gamma}) = \left\{
  \begin{array}{ll}
    \vspace{1.3mm}
    0, \,& \epsilon'_{\gamma} \leq \epsilon'_{\Delta}-\delta \\
    \vspace{1.3mm}
    \sigma_{\Delta}, \,& \epsilon'_{\Delta}-\delta < \epsilon'_{\gamma} \leq \epsilon'_{\Delta}+\delta \\
    0,  \,&  \epsilon'_{\gamma} \geq \epsilon'_{\Delta}+\delta
  \end{array}\right.
\end{equation}

\noindent where $\sigma_\Delta \approx 0.5$~mb is the peak value of the cross section ($\Delta$ resonance), 
which occurs at $\epsilon'_{\Delta}\approx 340$~MeV, and $\delta\approx 100$~MeV \cite{PDG01}. 
With inelasticities of pion production processes
occurring close to threshold typically taking values, $K_{p} \approx 0.2$,

\begin{equation}
  R(E_{p}) \sim 0.2 \, \sigma_{p\gamma} \int_{(\epsilon'_{\Delta}-\delta)/2\Gamma_{p}}^{(\epsilon'_{\Delta}+\delta)/2\Gamma_{p}} 
   \frac{dn_{\gamma}}{d\epsilon_{\gamma}}d\epsilon_{\gamma} \approx 0.2 \left[ \frac{l_0e^{x}}{(1-e^{-x})} \right]^{-1}\,
\label{attenuation}
\end{equation}

\noindent where $l_0 = 5$~Mpc and $x = 10^{20.53} \mbox{~eV}/E_{p}$ (note- this calculation assumes
only the presence of the CMB photon target exists).

The rate curves for these processes, shown in the left-panel of fig.~\ref{energyloss_rates}, 
may now be well understood, with their maximum values being of the order,

\begin{equation}
R_{\rm max}\approx 
\begin{cases}
\frac{m_{p}}{m_{e} n_{\rm CMB}\sigma_{p\gamma}}&\mbox{for pair production}\\
\frac{m_{p}}{m_{\pi} n_{\rm CMB}\sigma_{p\gamma}}&\mbox{for pion production}.
\end{cases}
\end{equation}

\section{1.4 Cosmic Ray Proton Propagation}
Once UHECRs have been accelerated and escaped their source region, 
their journey through extragalactic space begins. 
UHECR protons, of energies above 10$^{20}$~eV, are not expected to be deflected sufficiently by 
Galactic or extragalactic magnetic fields that their source directional information will be removed. 
Along with this, their propagation is limited through their inelastic collisions with cosmic background 
photons, attenuating their propagation to $\leq 30$~Mpc distance scales  \cite{Greisen:1966jv,Zatsepin:1966jv}
(see eqn~\ref{attenuation}), 
i.e. about the size of the local supercluster of galaxies.
Recent measurements by the Auger detector indicate that both that the arriving UHECR spectrum is 
not isotropic at high energies and that a suppression is observed in the arriving UHECR flux 
\cite{Abraham:2007si,Roth:2007in}.
However, more data is required for further clarity as to the cause of these results.
In preparation for developing such an understanding, we here outline the theoretical tools required
in order that calculations of UHECR proton propagation may be carried out. 

With very small inelasticities for the proton in each interaction, to a very good approximation pair 
production can be treated as a continuous energy loss process. Photo-pion production interactions, however,
involve very large inelasticities, so may not be treated as a continuous loss process and it is necessary 
to use Monte Carlo techniques \cite{Mucke:1999yb}.

It is also possible for UHECR protons to produce pions through interactions with the cosmic infrared background 
\cite{Hooper:2006tn}. Although the rates for these interactions are subdominant in comparison to energy losses 
from pair production, they can be important in determining the spectrum of cosmogenic UHE-neutrinos produced 
through UHECR proton propagation \cite{Stanev:2004kz}.

At present there are conflicting claims concerning the existence of the GZK cut-off, with two of the
three experiments with largest total exposure reporting a cut-off signature 
\cite{Takeda:1998ps,Abraham:2008ru,Abbasi:2007sv}.
Interestingly, departures from the ``vanilla-sky'' expectation for a GZK suppression are not so difficult
to obtain with both exotic new physics and less-exotic scenerios through alterations to some of the standard 
assumptions. Indeed, with regard these less exotic scenarios, the actual feature to be expected for the cut-off is 
dependent on several factors such as the local source distribution, composition, cross-over energy at
which the extragalactic UHECR flux dominates \cite{Taylor:2008jz}.
In particular, if UHECRs are 
not protons but consist of a substantial amount of heavy nuclei 
(see for example \cite{Anchordoqui:1999cu,Szabelski:2002rv}) then the 
spectrum will be altered from the usual expectation. Although evidence for the presence of heavy nuclei in UHECRs 
has been around for some time, the information available is still imprecise. 

\section{1.5 An Analytic Description for Proton Propagation}
For a source at distance $L$, the ratio of the number of protons, $N_n(E_p,L)$, 
arriving with energy $E_p$ having undergone $n$ pion production interactions to the initial population, 
$N_0(E,0)$, of protons of energy $E$ is \cite{Taylor:2008jz}

\begin{equation} \label{N_n}
  \frac{N_n(E_p,L)}{N_0(E,0)}=\sum_{m=0}^{n} l_0 \, l_m^{n-1} \exp(-L/l_m) 
  \prod_{p=0}^{n} \frac{1}{l_m - l_p}\,\,,
\end{equation}

\noindent with

$$l_m = \frac{l(E_p)}{(1-K_{p})^m}\, \hspace{4mm}\,\mbox{and} \hspace{4mm} 
l(E_p) = \frac{l_0}{e^{-x}(1-e^{-x})}\,\, ,$$

\noindent where $K_p$ is the inelasticity, $l_0=1$ Mpc and $x=10^{20.5}~{\rm eV}/E$. 

Thus, the distribution of protons of a given energy having not undergone any
pion production interactions is given by a simple exponential decay function. Similarly,
the distribution of mono-energetic protons having undergone one pion production interaction
but not having undergone a second pion production interaction is given by the difference
of two exponential functions. 

A demonstration of the success of this description may be found in \cite{Taylor:2008jz}, for which the
arriving UHECR flux from the local region is obtained and compared with the numerical result.
Interestingly, a similar expression as to that provided above (eqn \ref{N_n}) may also describe the
distribution of nuclear species produced following UHECR nuclei propagation, this will be discussed
further in section 1.7.

\section{1.6 Cosmic Ray Nuclei Propagation}
\begin{figure}
\centering\leavevmode
\includegraphics[width=2.3in,angle=-90]{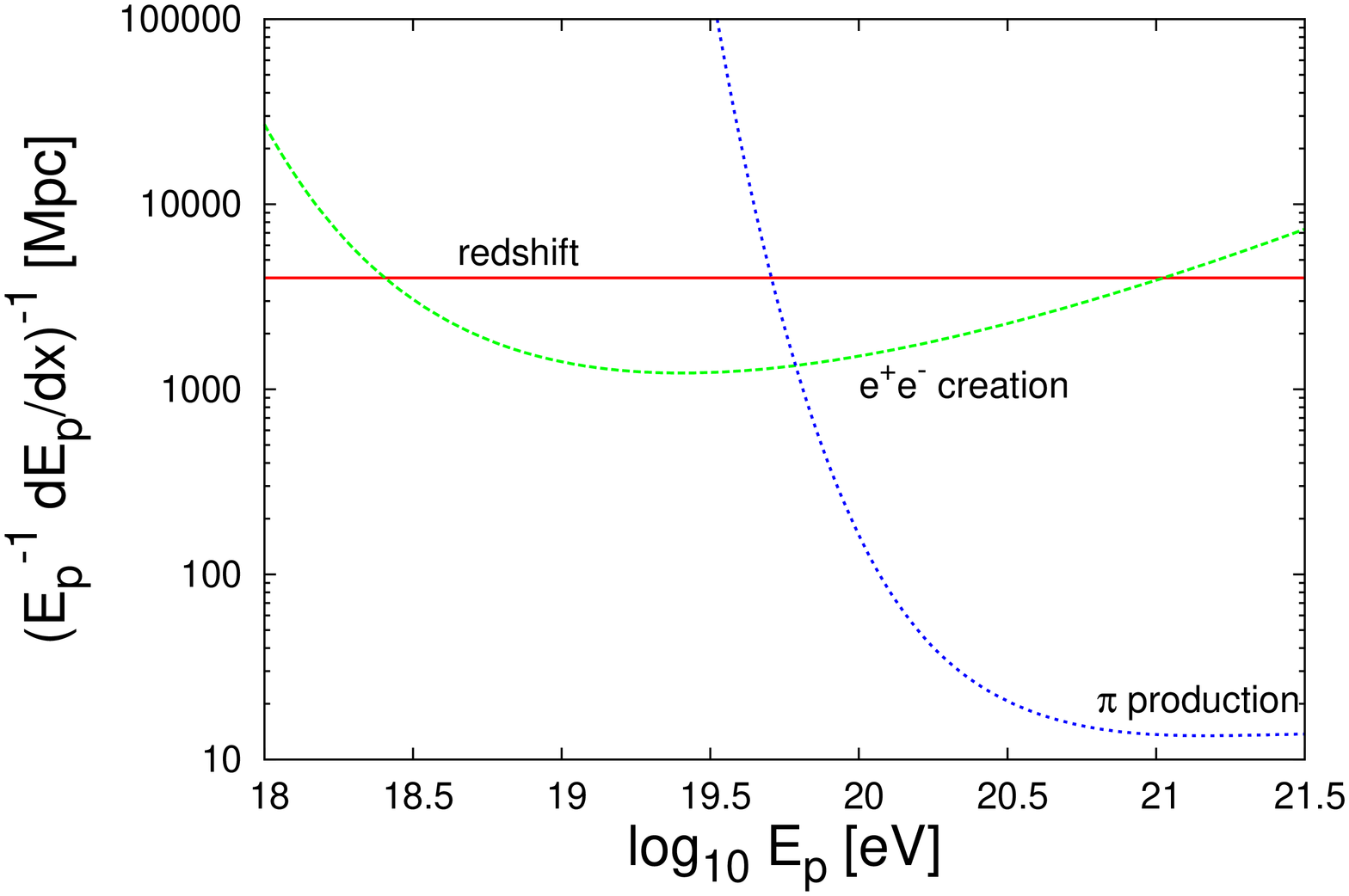}
\includegraphics[width=2.3in,angle=-90]{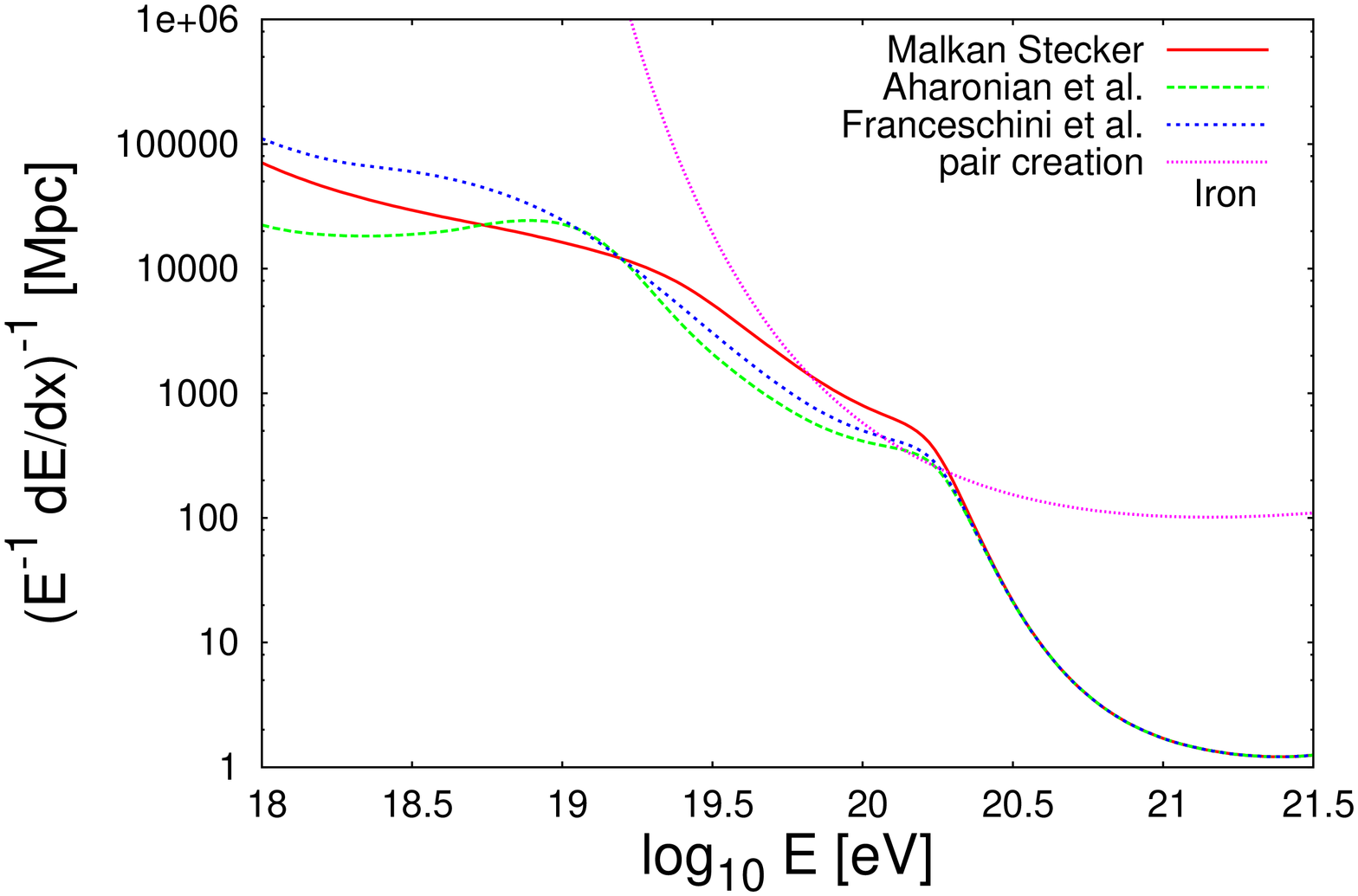}
\caption{{\bf Left-Panel:} Energy loss rates due to proton interactions with the CMB background photons. {\bf Right-Panel:} Energy loss rates due to nuclei interactions with CMB and CIB background photons.}
\label{energyloss_rates}
\end{figure}

In this section we study the intergalactic propagation of UHECR nuclei and their spectrum at Earth for different 
species injected at source. Nuclei undergo different interactions with the CMB/CIB 
\begin{equation}
(A,Z)\gamma\rightarrow
\begin{cases}
(A,Z)e^{+}e^{-}&\mbox{for}10^{19.7}{~\rm eV}<E_{(A,Z)}<10^{20.2}{~\rm eV}\cr
(A-1,Z)n/p&\mbox{for}E_{(A,Z)}<10^{19.7}{~\rm eV}\;\mbox{and}\;E_{(A,Z)}>10^{20.2}{~\rm eV}
\end{cases}
\end{equation}
Moreover, nuclei suffer photon-disintegration 
$$
(A,Z)\gamma\rightarrow (A^{\prime},Z^{\prime})+(Z-Z^{\prime})p+(A-A^{\prime}+Z-Z^{\prime})n
$$
at a rate
\begin{equation}
R=\frac{1}{2\Gamma_{(A,Z)}^{2}}\int_{0}^{\infty}\frac{1}{\epsilon_{\gamma}^{2}}\frac{dn_{\gamma}}{d \epsilon_{\gamma}}d\epsilon_{\gamma}\int_{0}^{2\Gamma_{(A,Z)}\epsilon_{\gamma}}\epsilon_{\gamma}^{\prime} \sigma_{A,Z,i_{p},i_n\gamma}(\epsilon_{\gamma}^{\prime}) K_{(A,Z)} d\epsilon_{\gamma}^{\prime}
\end{equation}
where $\Gamma_{(A,Z)}=E_{(A,Z)}/Am_{p}c^{2}$, $A$ and $Z$ denote the total number of nucleons (protons+neutrons) and charge of the nucleus, $i_{p}$ and 
$i_{n}$ are the number of protons and neutron broken off from a nucleus interaction and 
$\sigma_{A,Z,i_p,i_n\gamma}(\epsilon_{\gamma}^{\prime})$ is the cross-section (see \cite{Hooper:2006tn}). 
In the right-panel of Fig.~\ref{energyloss_rates} we present the energy loss lengths for iron nuclei
through pair creation photo-disintegration interactions. 

In general, during propagation, a heavy nuclei will undergo many photo-disintegration reactions, 
cascading down in atomic number and charge, 
generating secondary UHECR protons, neutrons and alpha particles along the way, each of which may also 
contribute to the UHECR spectrum at Earth. 
The amount of photo-disintegration depend on the amount of time they spend in the 
radiation fields, the strengths of the radiation fields that they propagate through, and the composition of the 
CRs leaving the accelerating region (source) \cite{Hooper:2006tn}.  

For the purpose of CR propagation simulations, for energies above $10^{19}$~eV, only 
sources up to redshift of $0.5$ really contribute to the CR flux at Earth, this limit being set by pair production 
energy losses of CRs during propagation. In general terms, during the past decade there has been quite a bit of 
work on the propagation of UHECR nuclei (see for example ref.s~\cite{Allard:2005ha,Hooper:2006tn}). 
Usually, the complex process of the 
photo-disintegration of UHECR nuclei into lighter nuclei and nucleons has been so far addressed using Monte Carlo 
techniques. More recently, a new analytic approach has been implemented with very good results 
\cite{Hooper:2008pm}. We will describe this subject briefly as a final point in our discussion.  

\section{1.7 An Analytic Description for Nuclei Propagation}
Denoting the number of nuclei with atomic 
number $A$ by $N_{A}$, the differential equation describing the population of a nuclei state is given by
\begin{equation}\label{Ngeneral}
\frac{dN_A}{dL}	+\frac{N_A}{l_{A\rightarrow A-1}}+\frac{N_A}{l_{A\rightarrow A-2}}+\ldots=
\frac{N_{A+1}}{l_{A+1\rightarrow A}}+\frac{N_{A+2}}{l_{A+2\rightarrow A}}+\ldots	
\end{equation} 
where $L$ is the distance traveled and $l_{i\rightarrow j}$ is the interaction length for state $i$ to disintegrate in 
state $j$. In order to simplify the calculations, we consider the case in which only a single nucleon loss process 
occur. This reduces the number of states in the system dramatically, along with the number of possible transitions. 
The equation \eqref{Ngeneral} simplifies to
\begin{equation}\label{Nsimple}
\frac{dN_A}{dL}	+\frac{N_A}{l_{A}}=
\frac{N_{A+1}}{l_{A+1}}	
\end{equation} 
The solution to this set of coupled differential equations, constrained by the initial conditions 
$N_n (L=0)\neq 0$ and $N_n (L=0)= 0$ for $A\neq n$ is given by
\begin{equation}\label{Nsol}
\frac{N_{A}(L)}{N_{n}(0)}= \sum^{n}_{m=A}l_{A}l_{m}^{n-A-1}\exp{(-\frac{L}{l_{m}})}\prod^{n}_{p=A(\neq m)}\frac{1}{l_{m}-l_{p}}	
\end{equation} 
A demonstration of the success of this description may be found in \cite{Hooper:2008pm}, for which the
arriving UHECR nuclei flux and composition is obtained and compared with the numerical result.

\section{1.8 Conclusion}
In this lecture, the dominant energy loss interactions of UHECR protons during their propagation 
through extragalactic radiation fields were described. 
An analytic description for proton propagation through such radiation fields was also provided.
Following a motivation for the consideration of UHECR nuclei propagation, from
measurements made by present generation UHECR detectors such as Auger, we also considered
the dominant energy loss in interactions of UHECR nuclei. An analytic description for this
situation was also provided.

\section{Lecture 2: The Cosmogenic UHE-Neutrino Flux}
During the propagation of ultra-high energy (UHE) protons through extragalactic space, a finite but
small probability exists that they may undergo an interaction with each of the background photons
they encounter. The most abundant of the background photons encountered, with sufficient energy to
partake in these interactions, existing in the 2.7~K microwave (10$^{-3}$~eV) and infrared (10$^{-2}$~eV) 
backgrounds. The large distances present in extragalactic space, however, provides sufficient
column densities of such photons between the source and Earth to make such interactions likely, 
leading to the limitation of UHECR proton propagation in extragalactic space 
(assuming rectilinear propagation) to distances less than $\frac{l_{0}e^{x}}{(1-e^{-x})}$, 
where $x=10^{20.53}~{\rm eV}/E_{p}$, $E_{p}$ is the proton's energy, 
and $l_{0}$ is 5~Mpc. The energy fluxes of secondaries produced through such interactions 
($p+\gamma\rightarrow p/n+\pi^{0}/\pi^{+}$) 
leads to the generation of both neutrino and photon fluxes. However, we here will focus on the production 
of the (cosmogenic) UHE-neutrinos, leaving the photon flux for the following lecture.

In the next section we will focus on the photo-pion producing interactions. This 
will be followed by a section describing the expected (cosmogenic) UHE-neutrino flux produced by
UHECR propagation through the cosmic background radiation fields, indicating how it is dependent
on the CR composition at the source.

\section{2.1 Source Distributions}
In order to calculate the cosmogenic UHE-neutrino spectrum arriving at Earth
it is necessary to assume an UHECR energy spectrum injected at source, 
$\left.dN_{p}/dE_{p}\right|_{\rm source}$, 
and the distribution of sources, described by the number of sources in shells of
redshift bins redshift surrounding the Earth, $dN_{\rm source}/dz$.

The energy spectrum assumed to be produced by our sources are motivated by Fermi
first order non relativistic shock acceleration theory.
In this theory, particles gain energy through consecutive crossings of a non-relativistic shock wave. 
The energy gain per crossing is found to be $\Delta E = \frac{4}{3} \beta E_{0}$, where $\beta$
is the relative velocity of the two plasma's either side of the shock in units of $c$ and $E_{0}$ is the 
initial energy of the particle. At the same time as the energy of the particles increases with each
crossing, the number of particles taking part is expected to decrease, with particles being lost from the 
system through advection (being washed downstream).
The number of particles lost in each crossing being given by,  $\Delta N = - \frac{4}{3} \beta N_{0}$, 
where $N_{p,0}$ is the initial number of particles in the system. 
After $n$ crossings, the energy of the particles will be 
$E=(1+\frac{4}{3}\beta)^n E_{0}$, whereas the number of particles will be 
$N=(1-\frac{4}{3}\beta)^n N_{0}$. From these distributions, one sees that of order $n \sim 1/\beta$ 
encounters are needed to significantly modify 
the particle energy and the number of particles. For $\beta \ll 1$, the altered particle distribution 
becomes $dN/dE \propto E^{-2}$, the spectrum of Fermi accelerated particles, with spectral index $\alpha = 2$.
Such a theoretical $E^{-2}$ spectrum is consistent with the inferred electron/proton spectra from $\gamma$-ray
observations in the GeV-TeV domain, produced by CR electron/proton energy loss processes close to the source 
\cite{Gralewicz:1997,Aharonian:2000iz,Aharonian:2006au}.

From this motivation, we assume a UHECR spectrum at the source described by a power law with an 
exponential cutoff $\left.dN/dE\right|_{\rm source} \propto E^{-\alpha} \,e^{-E/E_{\rm max}}$, where the maximum 
energy is\footnote{following Hillas criterion type arguments, the maximum energy
attainable by a source is expected to be proportional to the charge $Z$ of the particle being
accelerated} $E_{\rm max}=(Z/26)\times 10^{22}$~eV. 

For the distribution of sources per co-moving volume the quasar luminosity density 
evolution\cite{Boyle:1997sm} is adopted

\begin{equation} \label{dn/dV}
  \frac{dN_{\rm source}}{dV} \propto  \left\{
  \begin{array}{ll}
    (1+z)^3, \,& z < 1.9 \\
    (1+1.9)^3, \,& 1.9 < z < 2.7 \\
    (1+1.9)^3 \,\exp [-(z-2.7)/2.7],  \,& z > 2.7 
  \end{array}\right.
\end{equation}

\section{2.2. Cosmogenic UHE-Neutrino Production}
Neutrinos created during UHECR propagation are known as cosmogenic UHE-neutrinos.
As discussed in the first lecture, after leaving the source UHECR protons
may interact with the CMB photons creating charged pions through photo-pion production
interactions (see section 1.3 in the first lecture).
Following their decay, charged pions produce a flux of neutrinos,

\begin{equation} \label{pi_production}
  \pi^+ \rightarrow \nu_\mu + \mu^{+} \rightarrow \nu_\mu + e^+ +  \nu_e + \bar{\nu}_\mu \,.
\end{equation}

\noindent Neutrinos are also produced via beta decay of secondary neutrons

\begin{equation} \label{n_decay}
  n \rightarrow \, p + e^- + \bar{\nu}_e\,.
\end{equation}

\noindent The energy of the neutrinos resulting from $p\gamma$ reactions is 
$E_\nu \sim 0.05 E_p$, and in the case of neutron decay,  $E_\nu \sim 0.0005 E_p$. 

\begin{figure}

\includegraphics[width=2.5in,angle=0]{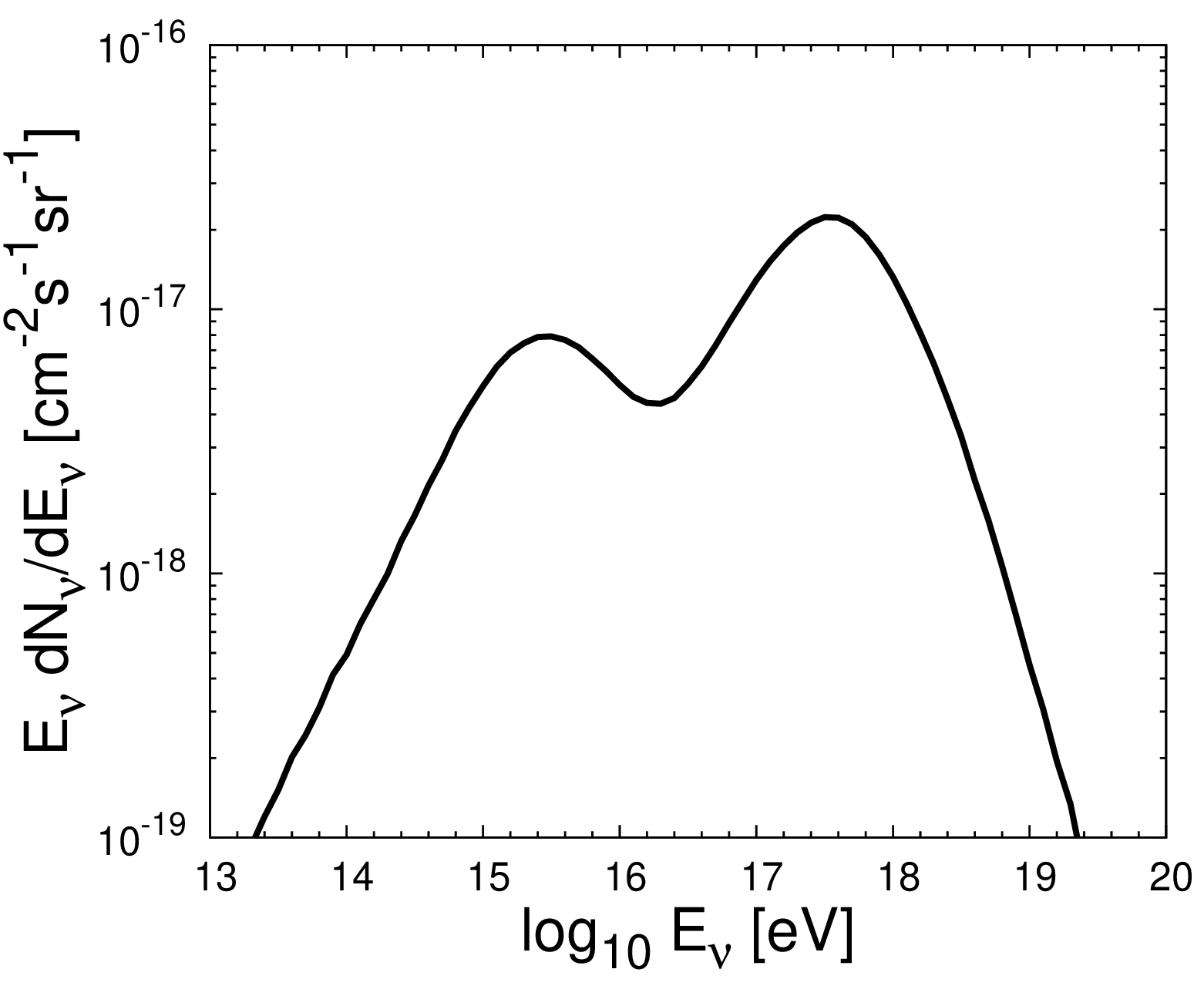}
\includegraphics[width=2.5in,angle=0]{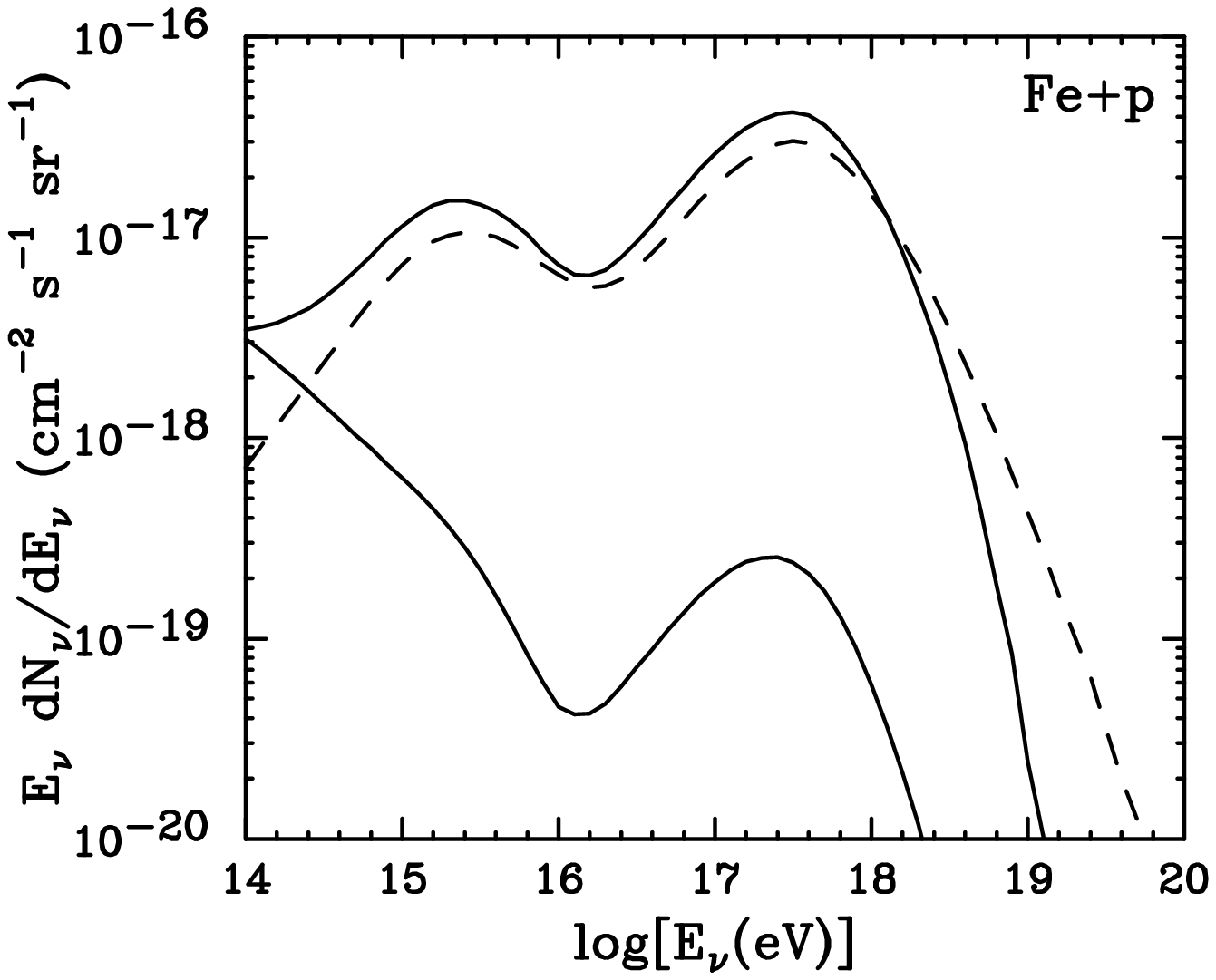}
\caption{{\bf Left-Panel:} The calculated cosmogenic UHE-neutrino flux produced by the cosmic population of UHECR protons assuming the energy spectrum and spatial distribution given in section 2.1. This result sits in good agreement with that calculated in ref.~\cite{Engel:2001hd}. {\bf Right-Panel:} The calculated cosmogenic UHE-neutrino flux produced by a cosmic population of UHECR protons and nuclei assuming the energy spectrum and spatial distribution of sources given in section 2.1. This result was obtained in ref.~\cite{Anchordoqui:2007fi}.}
\label{neutrino_fluxes}
\end{figure}

Following the assumption about the energy and spatial distributions of the sources, discussed in 
the previous section, we here obtain the subsequent cosmogenic UHE-neutrino flux produced.
In Fig.~\ref{neutrino_fluxes}, the charged pion decay generated neutrino flux at its maximum 
(the higher energy peak) 
is roughly three times the value of the neutron decay generated electron neutrino flux 
(lower energy peak). The normalisation of the cosmogenic UHE-neutrino energy flux, with a value
$\sim 10$~eV~cm$^{-2}$~s$^{-1}$~sr$^{-1}$, is comparable to the UHECR energy flux at 
$10^{19}$~eV, as might be expected by the large decrease in the attenuation length of UHECR protons 
at slightly higher energies, whose energy flux feeds into the UHE-neutrino flux.

Most models for the production of cosmogenic UHE-neutrinos assume that UHECRs are protons. 
This leads to an observable flux of neutrinos in detectors such as IceCube or KM3net. 
However, the existence of arriving UHECR nuclei can imply that these cosmogenic 
UHE-neutrino fluxes are suppressed by factors as large as $\sim 100$ (see \cite{Anchordoqui:2007fi}), 
making it hard to detect the (previously considered) ``guaranteed'' flux of cosmogenic UHE-neutrinos. 
Instead, we here calculate also the neutrino flux expected for the case of UHECR nuclei.

During propagation, UHECR nuclei experience photo-disintegration into their constituent 
nucleons in scattering off the CMB and CIB photons \cite{Hooper:2006tn}. The interaction of the 
daughter nucleons with these background photons may create charged pions which will go on to decay 
producing a cosmogenic UHE-flux of neutrinos. 
In the nuclei rest frame, the photo-disintegration cross section peaks at a photon energy of 
$\sim 30$ MeV (giant dipole resonance). Since in the lab frame, the average CMB-CIB photon 
energy is $\sim 10^{-2}$~eV, the nucleus must have an energy of approximately $10^{20}$~eV 
for the interaction to take place. This energy is roughly the same as that needed 
for protons to interact with the CMB photons. Thus, the GZK suppression for nuclei can be 
expected to occur at similar energies to that for protons.

For a given spectrum and composition at the sources, the calculated spectrum at Earth must be 
consistent both with the observed UHECR spectrum and the 
$X_{\rm max}$ (the atmospheric depth of the shower maximum) data.
Whereas, the $X_{\rm max}$ data seems to find better agreement for an all-iron spectrum 
compared to an all-proton composition at source (see for instance \cite{Anchordoqui:2007fi}),
the UHECR spectrum does not give a clear indication.

The cosmogenic UHE-neutrino flux for CR nuclei is presented in the right-panel of 
Fig.~\ref{neutrino_fluxes}, together with the all-proton ``guaranteed'' neutrino flux (dashed line).
 The upper and lower solid line corresponds, respectively, to the highest 
(dominant proton composition) and lowest (dominant iron admixture) neutrino flux compatible 
with the data. It can be seen that the lowest possible value is two orders of magnitude
 below the previously considered all-proton cosmogenic UHE-neutrino flux. If this is the case, 
it will not be possible to observe the cosmogenic UHE-neutrinos in detectors such as IceCube or KM3net.

\section {2.3 Conclusions}
In this lecture, a motivation for injection spectrum spectral index close to 2 was provided through the
consideration of the Fermi first order non-relativistic shock acceleration mechanism. Through simulations
of sources with such spectra, with an assumed cosmological evolution history, the diffuse cosmogenic UHE-neutrino
flux was obtained. Further to this, with the Auger $X_{\rm max}$ data presently suggesting that UHECRs inject a 
component of heavy nuclei at source, this diffuse cosmogenic UHE-neutrino flux calculation has also been
carried out for injection spectra and composition consistent with both the arriving spectrum and $X_{\rm max}$
Auger measurements. In this way, the cosmogenic UHE-neutrino flux calculation has been demonstrated to rest upon 
important underlying assumptions, with the flux typically determined being by no means guaranteed. 
The (cosmogenic) UHE-neutrino flux may be reduced up to a factor of 100 if instead of the all-proton 
(previously studied) scenario, a large heavy nuclei component at the sources exists\cite{Anchordoqui:2007fi}.

\section{Lecture 3: The UHECR Photon Fraction}

\section{3.1 UHE-Photon Production through UHECR Losses}
Following the acceleration of UHECR by their sources, and subsequent escape, this flux of particles 
(if protons) may be attenuated by photo-pion production 
interactions ($p+\gamma\rightarrow p/n+\pi^{0}/\pi^{+}$) with the background radiation photons. 
Through this attenuation, a subsequent secondary flux of UHE-photons is
generated via the decay of neutral pions ($\pi^{0}\rightarrow \gamma\gamma$).
From isospin considerations for the Delta-resonance decay, neutral pions 
are twice as likely to be produced as charged pions. However, close to threshold ($\epsilon_{\gamma}'\approx 145$~MeV) 
where direct pion production dominates, charge pion production is more likely. Nevertheless, unlike high energy 
neutrinos produced through the decay of charged pions 
($\pi^{+}/\pi^{-} \rightarrow \mu^{+}/\mu^{-}+\nu_{\mu}/\bar{\nu}_{\mu}\rightarrow e^{+}/e^{-}+\nu_{\mu}+\bar{\nu}_{\mu}+\nu_{e}/\bar{\nu}_{e}$), 
these UHE-photons are not free to propagate through the Universe over cosmological distances. 
The UHE-photon flux is in fact attenuated through interactions with the background radiation field via pair 
production $\gamma\gamma\rightarrow e^{+}e^{-}$ on Mpc size scales, similar to the scales from their sources on 
which these photon fluxes are produced.

UHE-electrons produced through such pair creation interactions may either subsequently undergo inverse Compton 
cooling interactions off the background radiation field through high center-of-mass $e\gamma$ collisions, 
leading to a repeating cycle of pair production and inverse Compton scattering and the development of an 
electromagnetic cascade \cite{Protheroe1,Protheroe2, Aharonian:1992qf}, or synchrotron cool in the extra-galactic 
magnetic fields \cite{Aharonian:1992qf, Gelmini:2005wu}. This is investigated in detail in the following section. 
To conclude this section we would like to remark that high energy photon production is an inevitable consequence of 
the GZK cut-off's existence.  

\section{3.2 The Propagation of UHE-Photons}
In this section we discuss how the UHE-photon flux, produced near the UHE-proton sources, is attenuated 
by their interaction with different cosmic background radiations and extragalactic magnetic fields.
UHE-photon interaction with background radiation fields may lead to an electromagnetic cascade generated 
before these generated UHE-photons reach Earth.
Such a cascade being generated by the repetition of two processes: (1) pair creation through UHE-photon interactions
with background photons and (2) photon production by UHE-electrons inverse Compton scattering
background photons. However, with regards the second process, 
the UHE-electrons may also synchrotron cool in the extragalactic magnetic field.
In the following, we shall look more closely into the physics of pair creation and investigate where the 
energy content of the electrons and positrons produced flows to next and how the dominant cooling channel
is decided upon.

\begin{figure}
\includegraphics[width=2.3in,angle=-90]{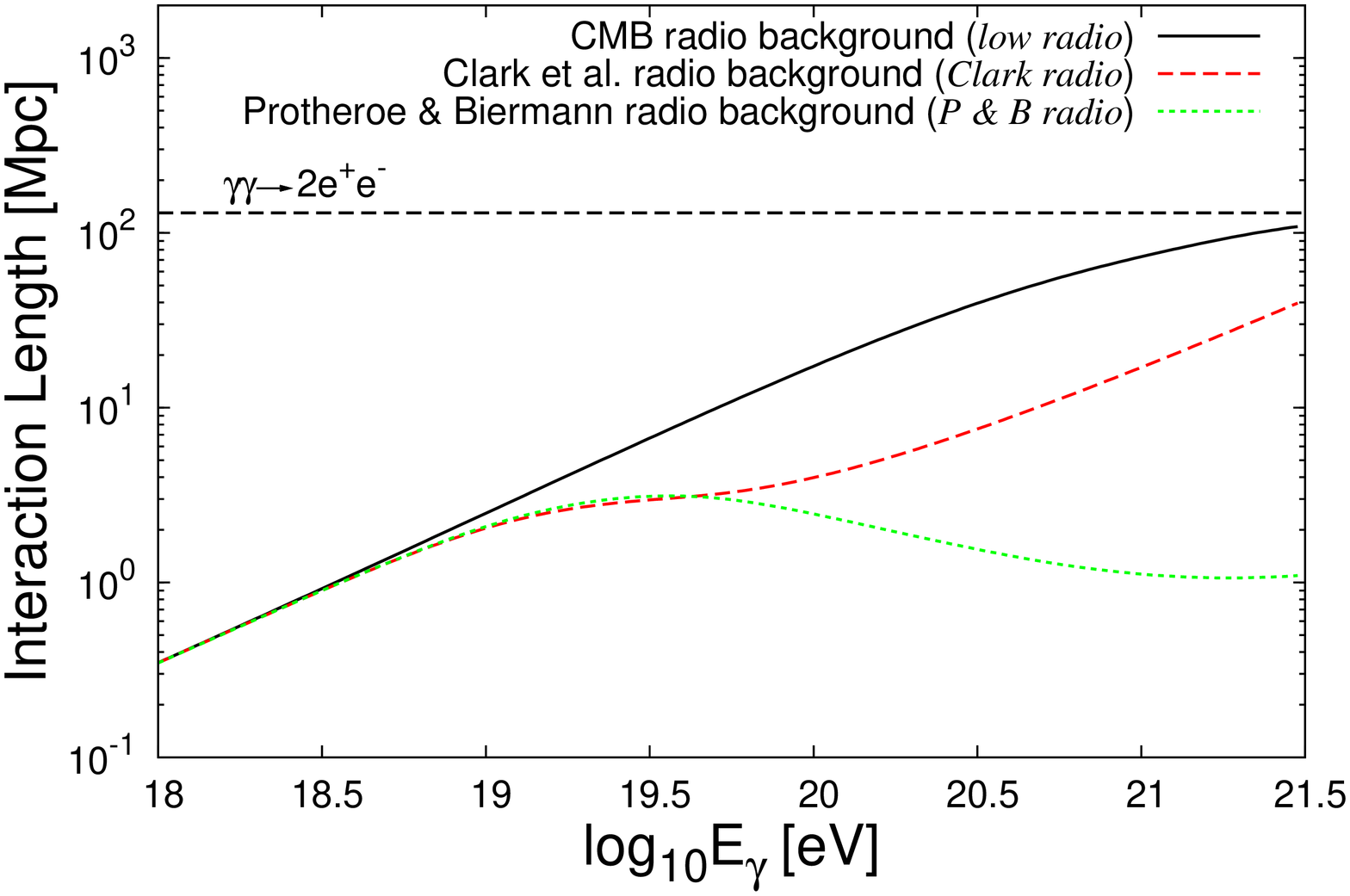}
\includegraphics[width=2.3in,angle=-90]{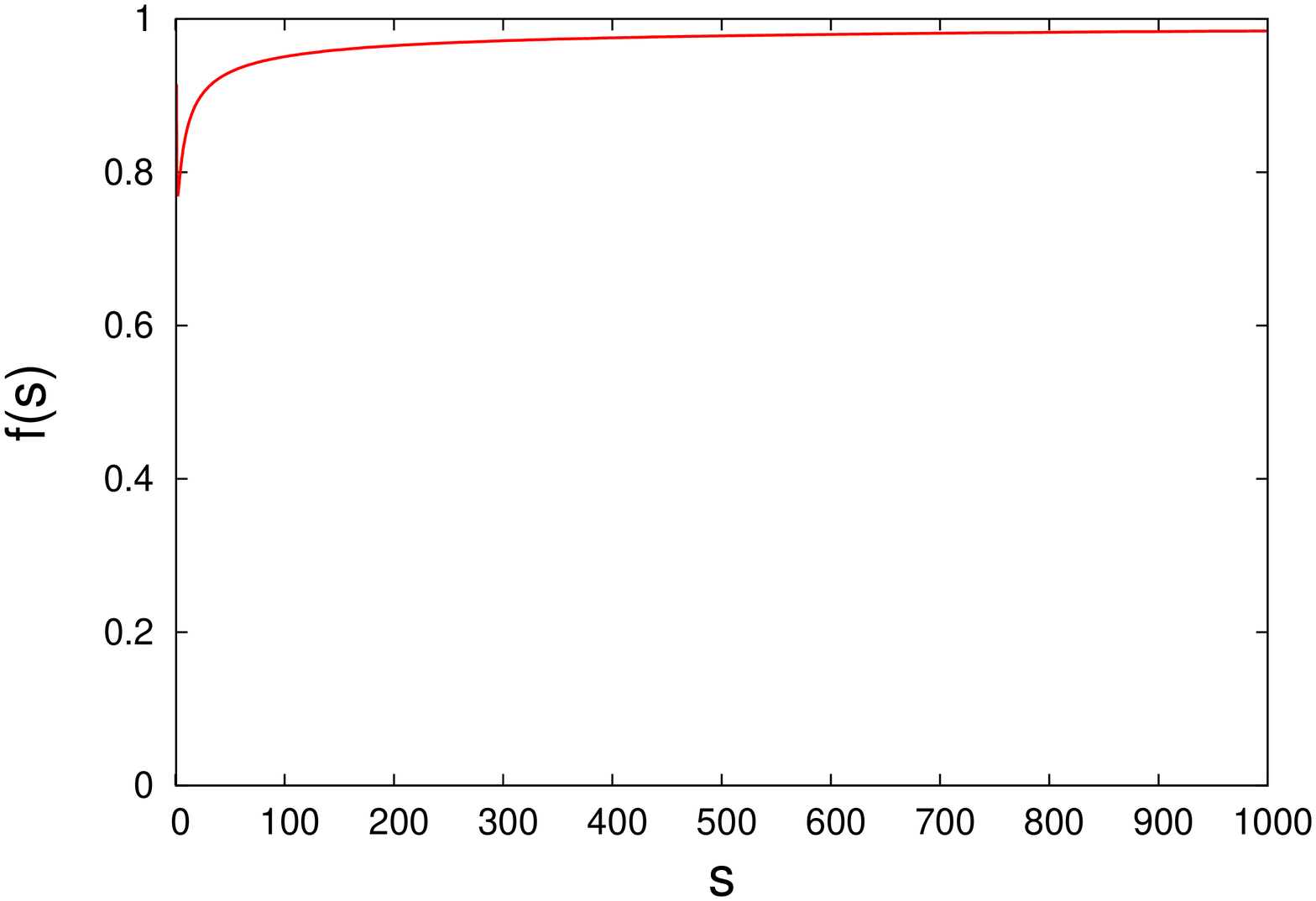}
\caption{{\bf Left-Panel:} The interaction lengths of UHE-photons propagating through the background radiation fields in extragalactic space. {\bf Right-Panel:} The function $f(s)$ for $\gamma\gamma$ interactions for an isotropic distribution of target photon momenta in the lab frame.}
\label{photon_interaction}
\end{figure}

In the center-of-mass frame the electron/positron pairs are produced with equal energy. 
However, following a boost back to the lab frame, one of the electron's tends to take nearly 
all the energy. The average ratio between the interacting particle $E_{e}$ and the particle produced 
$E_{\gamma}$, for $s\gg 1$, being approximately,
\begin{equation}
\frac{E_{e}}{E_{\gamma}}\approx\frac{4s-1}{4s}f(s)
\end{equation} 
where $s$ is the squared center-of-mass energy of the scattering process in units of $(2m_{e}c^{2})^{2}$ 
(for convenience since this sets $s=1$ as threshold for the process), given by
\begin{equation}
s=\frac{4E_{\gamma}E_{\gamma}^{\rm bg}}{(2m_{e} c^2)^2}
\end{equation}
for a head-on collision in the lab frame. A description of $f(s)$ is shown in the right-panel of
Fig.~\ref{photon_interaction}.

For large $s$, one of the electrons produced carries away most of the original photon energy. 
What happens with this electron energy? The UHE-electrons may lose energy by the interaction either with the 
CMB radiation field (through Inverse Compton scattering) or with the extragalactic magnetic field 
(through the synchrotron channel). Since the energy density of both fields are 
$U^{\rm CMB}_{\gamma}=0.25$~eV~cm$^{-3}$ and $U_B=10^{-8}$~eV~cm$^{-3}$ ($B=3\times 10^{-10}$~G),
one might naively expect that the dominating interaction will be the field with the larger
energy density (ie. the inverse Compton process).
However, we here pay a closer look at the physics of these competing cooling channels.

In the Thomson approximation, the center-of-mass frame is the electron's rest frame, so the 
background photon in the electron's frame simply has its momentum reversed during the interaction. This occurs 
when $\Gamma_{e} E_{\gamma}^{\rm bg}<m_{e} c^2$, where $\Gamma_{e}$ is the electron's Lorentz factor and $m_{e}$ is its
rest mass. For synchrotron photons, we consider the magnetic field as a virtual field of photons, with
characteristic energies,
\begin{equation}
E_{\gamma}^{\rm bg}=\left(\frac{B}{B_{\rm crit.}}\right)m_{e}c^2, 
\end{equation}      
where $B_{\rm crit.}=4\times 10^{13}$~G. Hence, with $B\approx$nG a value for the virtual photons of $E_{\gamma}^{\rm bg}\approx10^{-18}$~eV is obtained. 
For $E_{e}=10^{19}$~eV ($\Gamma_{e}=2\times 10^{13}$), the boosted virtual photon's energy in the electron's rest frame is
$\Gamma_{e} E_{\gamma}^{\rm bg}\approx 2\times 10^{-6}$~eV (ie. well within the Thomson regime).

On the other hand, for the CMB photons with $E_{\gamma}^{\rm CMB}\sim 10^{-3}$~eV, such photons in the electron's rest frame
(for the same energy electron) have an energy $\Gamma_{e} E_{\gamma}^{\rm bg}=2\times 10^{10}$~eV, thus the Thomson approximation is not
applicable so the Klein-Nishina description (applicable for $\Gamma_{e} E_{\gamma}^{\rm bg}\geq m_{e} c^2$) must be applied instead. 
Once in the Klein-Nishina regime, the center-of-mass frame can no longer be taken as the electron rest frame and 
the inverse Compton scattering cross-section is suppressed, leaving the dominant cooling channel less obvious. 
If the CMB is the only background radiation field and the extragalactic magnetic field is roughly nG 
in magnitude, the magnetic field (synchrotron) cooling channel wins and UHE-photons will undergo cooling
interactions predominantly through the synchrotron channel. 

However, if there exists a stronger cosmic radio background (CRB) component ($E^{\rm bg}_{\gamma}\approx 10^{-6}$~eV) with $U_{\gamma}^{\rm CRB}\approx 10^{-8}$~eV~cm$^{-3}$ then 
$\Gamma_{e} E_{\gamma}^{\rm bg}=2\times 10^{5}$~eV thus the Thomson approximation still roughly holds and the 
inverse Compton channel can dominate. 
Under this condition, along with extragalactic magnetic fields of $\sim 10^{-2}$~nG, the inverse Compton channel 
will dominate the electron/positron cooling.      
In the scenario with dominant inverse Compton cooling, an $e/\gamma$ cascade 
is produced. If this cascade developes with interactions occurring in the Klein-Nishina regime, the high energy 
particle simply changes from a neutral to charge state and back spending roughly 1/3 of it's time in the 
neutral (photon) state and 2/3 in the charged (electron/positron) state.  

\section{3.3 The Proton/Photon Attenuation Rates and Ratio at Earth} 
\begin{figure}
\includegraphics[width=2.3in,angle=-90]{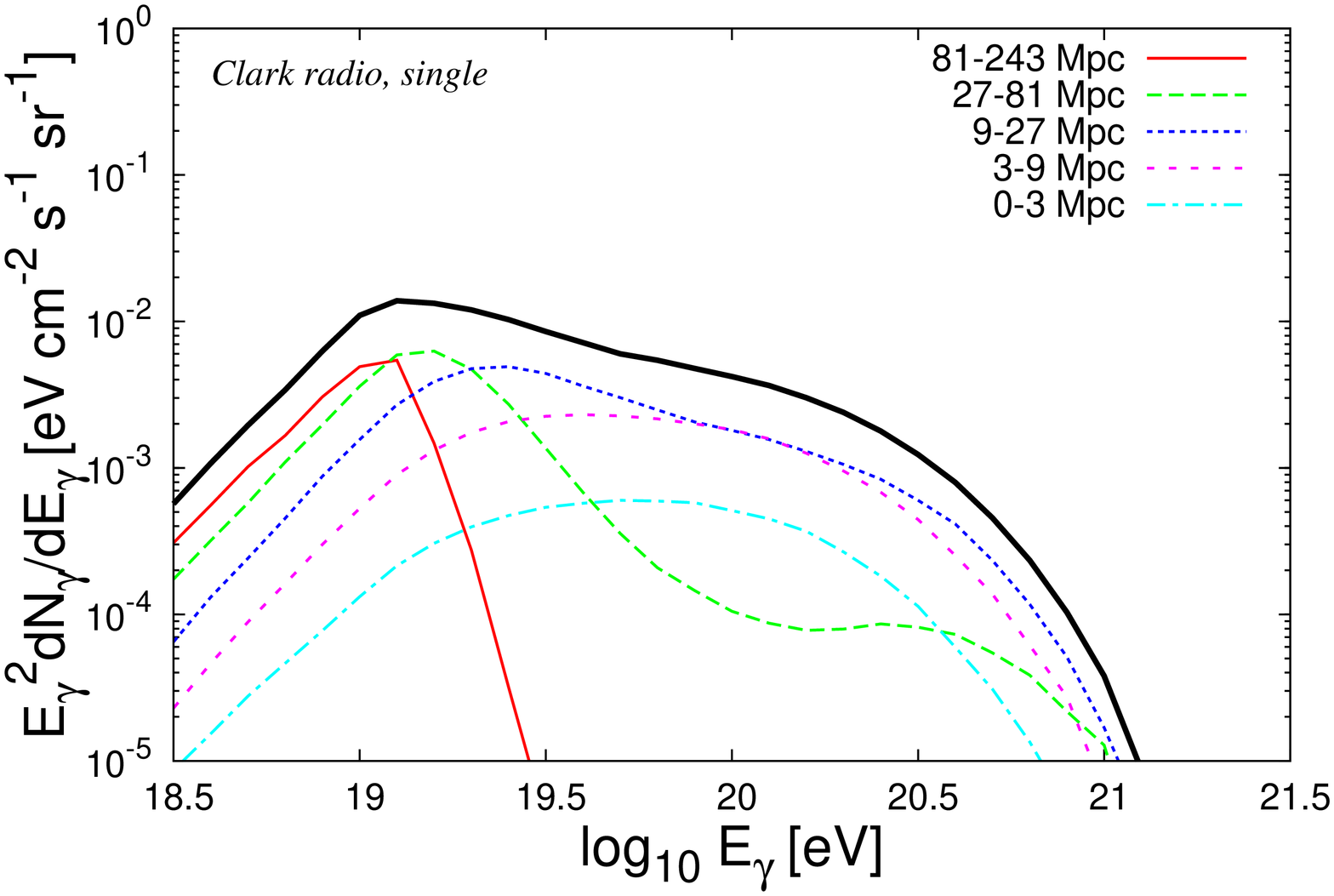}
\includegraphics[width=2.3in,angle=-90]{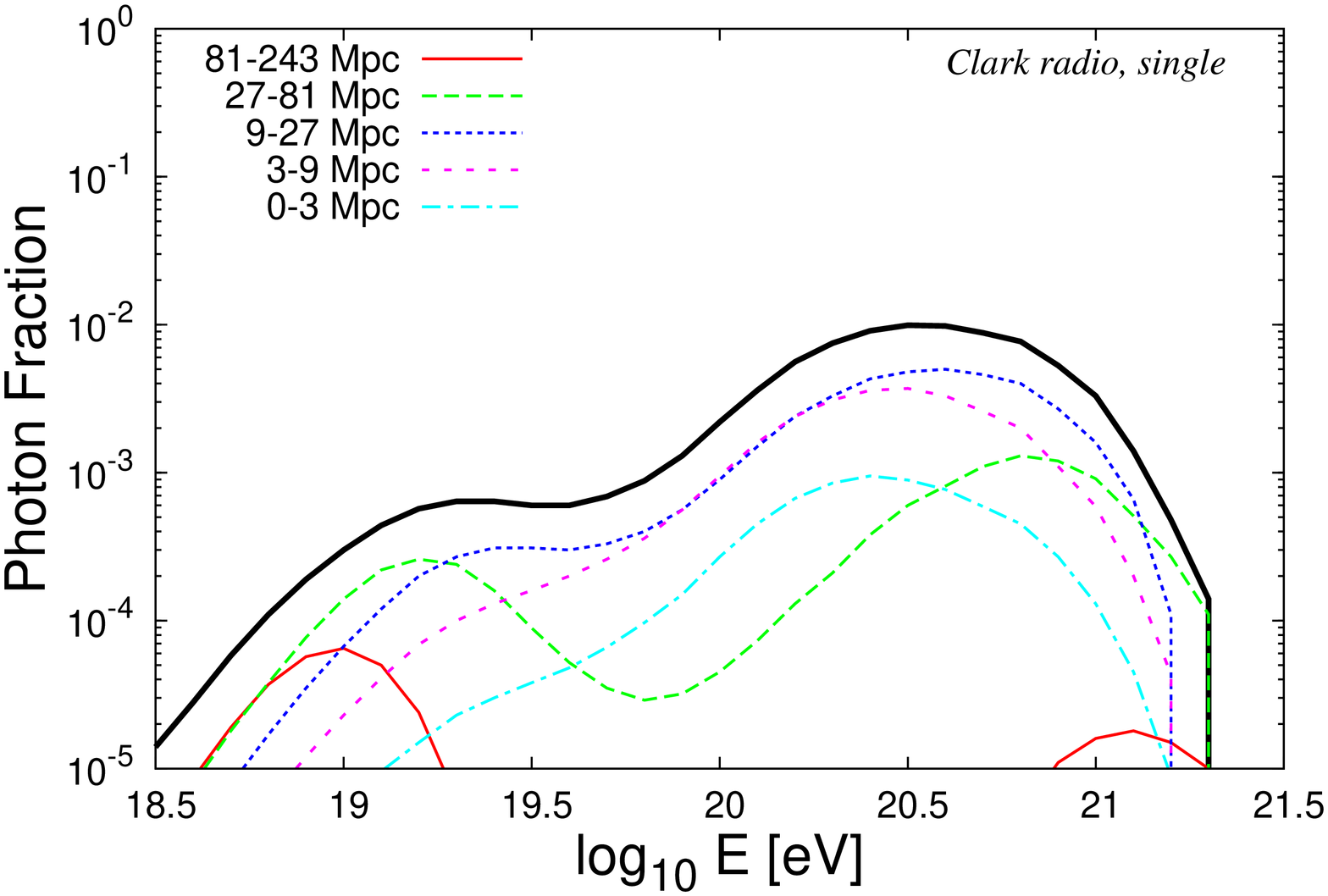}
\caption{{\bf Left-Panel:} The UHE-photon flux, produced through UHECR $\pi^{0}$ interactions, arriving at Earth, the different colour lines in the figure depicting a breakdown in the contributions to the photon flux from different shell radii.{\bf Right-Panel:} The photon fraction of UHECR arriving at Earth, the different colour lines in the figure depicting a breakdown in the contributions to the photon flux from different shell radii.}
\label{photon_results}
\end{figure}

In the last section we discussed the difficulties that a high energy photon undergoes in the Universe. 
We here look at a comparison of the proton and photon propagation through extragalactic space 
in order to study the photon/proton ratio expected at Earth. Considering UHE-photons as UHECR proton interaction 
products, it is necessary to calculate the UHE-photon interaction lengths with different possible realisations of
the background radiation field field, as shown in the left-panel of Fig.~\ref{photon_interaction}.

A clear difference in the interaction lengths of UHE-photons is seen to be expected for the Clark radio 
background value \cite{Parker:1970} and that provided purely by the radio component of the CMB, for UHE-photons with energies
$>10^{19}$~eV. Furthermore, a clear difference is seen to be expected between the interactions rates of 
UHE-photons in the Clark and Protheroe and Biermann \cite{Protheroe:1996si} backgrounds for energies $>10^{20}$~eV. 
However, for these calculations, the Clark radio background will be assumed.

With the physics of UHE-photon interactions discussed in the previous section, the question as to what fraction of
UHECR are photons may be addressed.
We here discuss the contribution from different source regions by separating out the fluxes 
produced from sources within the local $0-240$~Mpc region surrounding the Earth. We assume 
that UHECR sources have a homogeneous density distribution locally, with the number of sources scaling simply
with the co-moving volume element under consideration. 
Along with this, we assume each source is produces UHECR with a 
power law energy spectrum and exponential cut-off. 
These source spatial and energy distributions being given by
\begin{align}
\frac{dN}{dV_{c}}\propto (1+z)^{3}&&\frac{dN}{dE_{p}}\propto E_{p}^{-\alpha}e^{-E_{p}/E_{p}^{\rm max}}
\end{align}
with $\alpha=2$ and $E_{p}^{\rm max}=10^{20.5}$~eV. Following these assumptions, the arriving proton and photon fluxes were 
calculated for different sources shells of radii $0-3$~Mpc, $3-9$~Mpc, $9-27$~Mpc, $27-81$~Mpc, $81-243$~Mpc 
and energy range $10^{19}>E_{p}>10^{21}$~eV. The resulting arriving UHE-photon flux and photon fraction of UHECR
are shown in the left and right panels of Fig~\ref{photon_results}.

\begin{figure}
\includegraphics[width=2.3in,angle=-90]{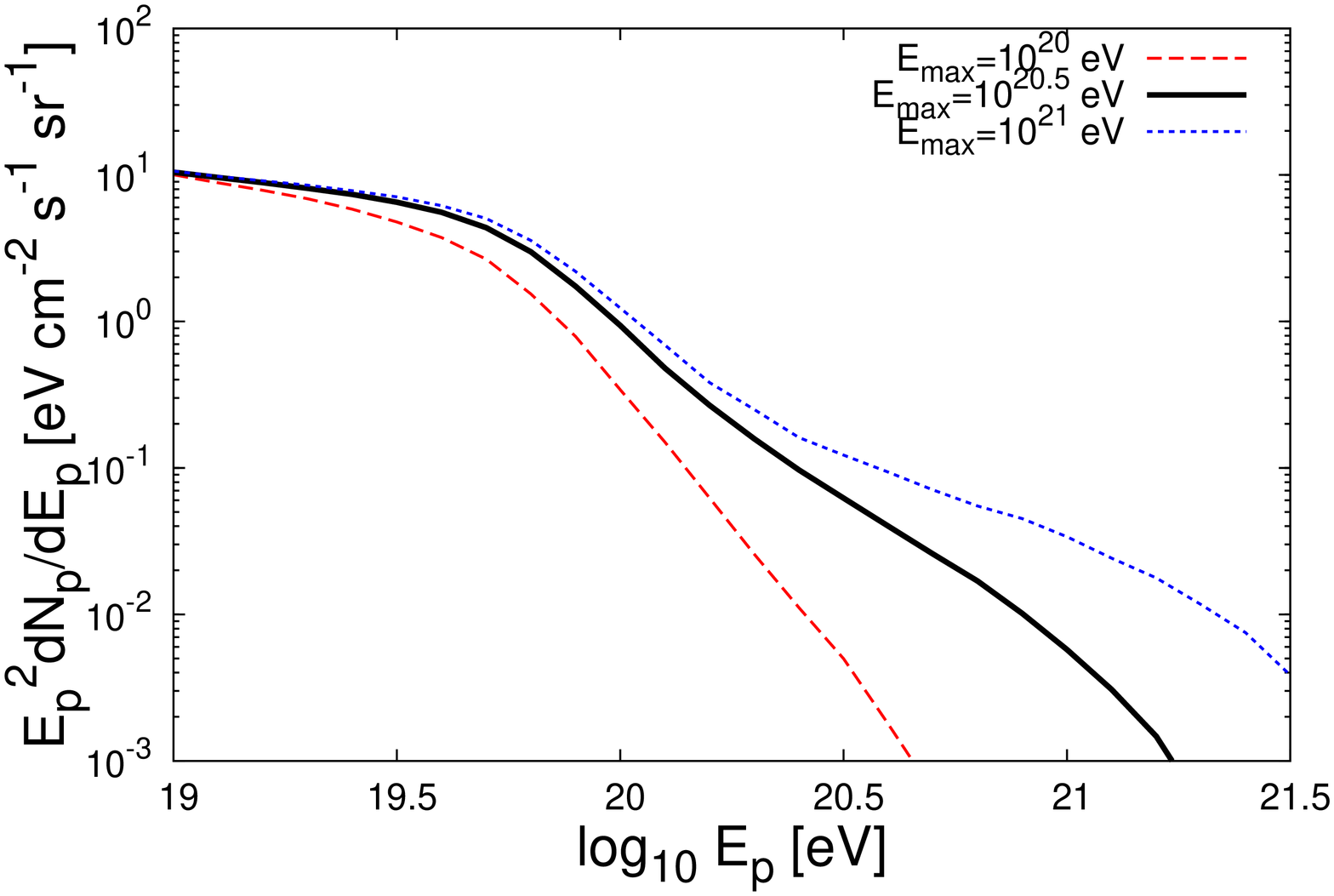}
\includegraphics[width=2.3in,angle=-90]{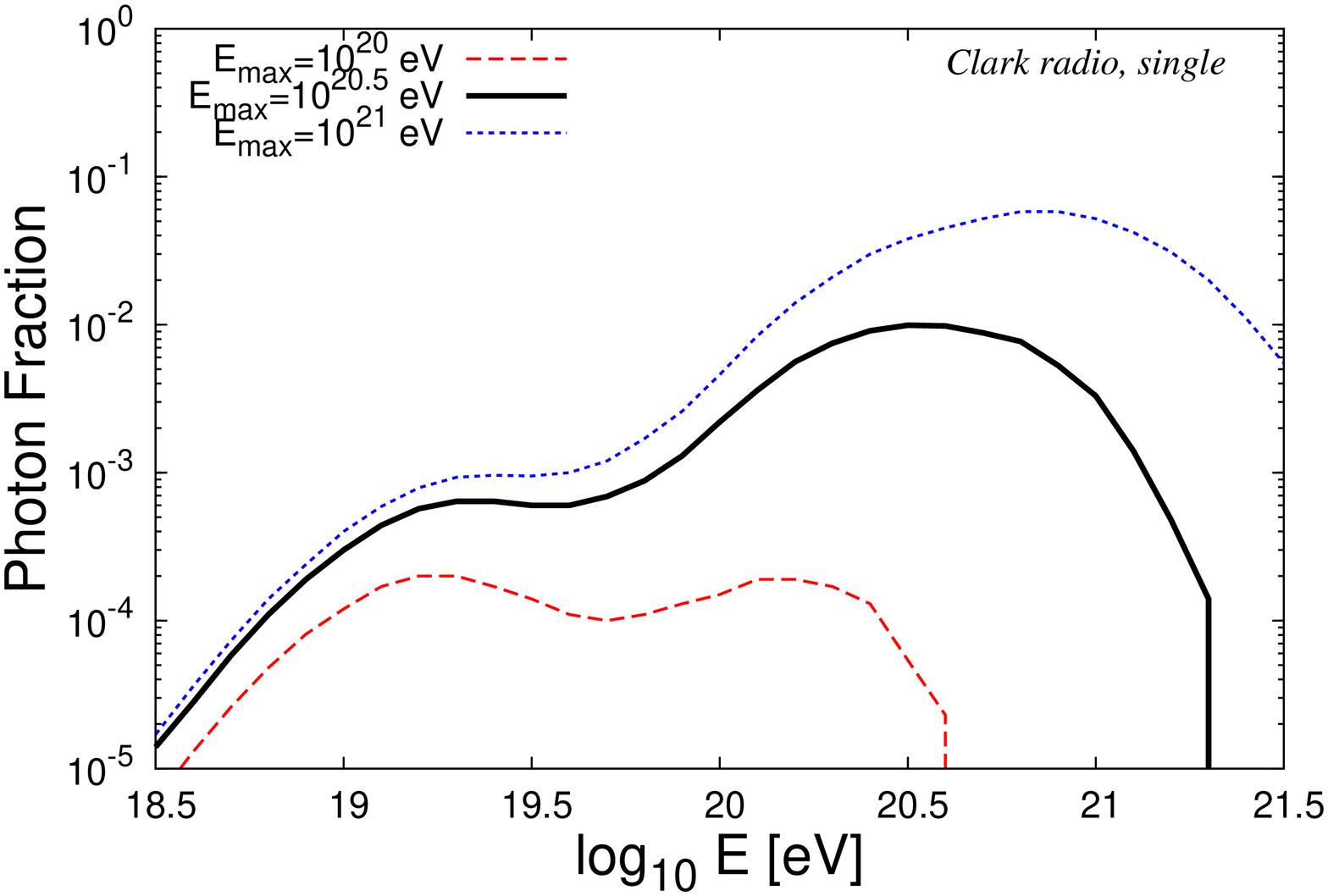}
\caption{{\bf Left}: The UHECR proton flux arriving from a homogeneous distribution of sources for different values of $E_{\rm max}$. {\bf Right}: The corresponding photon fraction of the arriving UHECR obtained from a homogeneous distribution of sources for different values of $E_{\rm max}$. For this plot, the label ``{\it Clark radio, single}'' refers to the assumptions about the radio background and magnetic field strengths, for which we have assumed the Clark et al. \cite{Parker:1970} radio background and extra-galactic magnetic fields stronger than 0.1~nG. The injected proton spectrum assumed for these results had a spectral index of $\alpha$=2.}
\label{Photon_Fraction_Cut-Off}
\end{figure}

With searches for UHE-photons in UHECR data providing photon fraction upper limits \cite{Risse:2007sd}, we
here look into the photon fraction values we expect through UHECR energy losses. Investigations into 
the photon flux produced through the GZK interactions have demonstrated that significant photon fractions can be 
expected for particular assumptions about the UHECR sources and extragalactic environment. In the 
``Clark radio, single'' scenario (assuming the Clark radio background and extragalactic magnetic field stronger 
than $0.1$~nG \cite{Parker:1970}) the arriving photon fraction is predominantly contributed from sources in relatively 
local shells, with distances tens of Mpc away. For the homogeneous source distribution case, photon fractions 
between $10^{-3}$ and $10^{-2}$ can be expected with the photon fraction peaking at $10^{-2}$, approximately the maximum energy the 
UHECRs are injected at ($E_{p}^{\rm max}=10^{20.5}$~eV). The effect of changing this maximum
energy are shown in Fig.~\ref{Photon_Fraction_Cut-Off} of the last lecture. Details about photon fraction can be 
found in \cite{Taylor:2008jz,Gelmini:2005wu,Kalashev:2007sn} 

\section{3.4 Conclusion}          
In this lecture, the energy losses of UHE-photons in both magnetic fields and radiation fields
have been discussed. 
We have demonstrated that this photon fraction provides a complementary measurement to that of the cut-off 
feature in the UHECR spectrum, allowing differentiation between the "tired source" and the GZK cut-off 
scenarios as well as carrying information about the energy injection spectrum at the source. 
Though measurements by Auger of the arriving UHECR flux photon fraction are only able to
presently provide lower limit values ($\sim 10^{-2}$), even smaller fractions should be probed
in the near future \cite{Aglietta:2007yx}. 
With energy loss scales $10-1000$~Mpc for protons and $1-100$~Mpc for photons both at $>10^{18}$~eV, and 
source distance scales expected $\sim 50$~Mpc, the UHECR proton and photon flux collectively are perfectly suited 
to use as a probe of the local distribution of UHECR sources (discussed in the following lecture). 

\section{Lecture 4: Multi-Messenger Approaches to Problems}
Protons, nuclei, gamma-rays and neutrinos produced in extragalactic UHECR sources, such as
Active Galactic Nuclei (AGN), Gamma Ray Bursts (GRBs) and Starburst Galaxies,
will encounter cosmic background (CB) photons during propagation through extragalactic space. While UHE-neutrinos 
travel essentially without losing energy, 
the other UHE-particles may interact with the CB photons, limiting their propagation range
to multi-Mpc scales:
$\sim$100~Mpc for protons; $\sim$100~Mpc for nuclei; and $\sim$10~Mpc for photons. 
This suggests the possibility of exploring cosmologically nearby sources ($\sim$50~Mpc) using 
UHECR protons, nuclei, photons and neutrinos simultaneously. 

This lecture is divided into two sections dealing with two different multi-messenger
approaches. In the first of these sections, the proton-photon connection looks into the 
employment of the UHECR spectrum and its photon fraction to ascertain from future UHECR flux measurements
the local distribution of UHECR sources. In this section, the arriving UHECR proton spectrum for different 
shells from a homogeneous distribution of sources is calculated and the photon fraction of UHECRs is 
obtained. The effects on these results due to a local source over/under-density are highlighted.
In the second section, we focus on the UHE-neutrino flux produced directly by the UHECR source for 
the candidate sources of AGN, GRB and Starburst Galaxies. 
We review the main features of the candidate UHECR sources (GRBs, AGN, starburst galaxies). 
Through a consideration of the source opacity, dictating the relative energy losses of
nucleons and nuclei before leaving the source region, the observation of a nuclei
component in the arriving UHECR at Earth is used to provide strong constraints on the UHE-neutrino flux expected 
from the whole ensemble of UHECR sources.
The opacity factor and neutrino production in these sources are discussed and 
the interaction rates within these sources obtained. We end the lecture by presenting our 
conclusions for the multi-messenger approaches discussed.

\section{4.1 UHECRs and UHE-photon fraction}
As discussed in the first lecture, UHECR protons during propagation mainly lose their energy 
through interactions with the CMB via pair creation and photo-pion production interactions.
Thus, as discussed in the third lecture, an associated UHE-photon flux arises from the decay 
of the neutral pions, $\pi^0 \rightarrow \gamma\gamma$, created through the 
$p+\gamma\rightarrow p+\pi^0$ interaction channel. 
These high energy photons also have their propagation range limited by background radiation fields (CMB and radio background). 
If, besides the CMB, a low radio background contribution is assumed, the interaction length for 
$10^{20}$~eV photons is approximately 10~Mpc (as shown in the left-panel of Fig.~\ref{photon_results}). 

To calculate the UHECR flux at Earth, a CR source spectrum of the form 
$dN_{CR}/dE_{CR} \sim E_{CR}^{-\alpha} \exp(-E_{CR}/E_{\rm max})$ is adopted, 
with $E_{\rm max}=10^{20.5}$~eV, the maximum energy CR energy attainable by the source. 
The value of the spectral index, $\alpha=2$, is motivated by non-relativistic first order Fermi acceleration 
(see first lecture).
The assumed spatial distribution of CR sources is $dN/dV \sim (1+z)^3$, where $z$ is the redshift of
the source. Note that for the cosmologically nearby sources, any $z$ dependence is of little relevance.

To describe the contribution from regions at different distances to the UHECR flux (ie. to the GZK feature), 
five shells are taken. If $L$ denotes a bracketed range of distances of the source regions from the Earth (in Mpc) 
and $S_{n}$ stands for shell $n$, we have: 
$S_1$ for $L\in [0,3]$, $S_2$ for $L\in [3,9]$, $S_3$ for $L\in [9,27]$, $S_4$ for $L\in [27,81]$, and $S_5$ for $L\in [81,243]$.%

For a homogeneous distribution of sources, the arriving contribution from accelerators within a shell,
ignoring the effects of attenuation, is proportional to the width of the shell. So the ratio of 
contributions from consecutive shells $S_{n+1}$ to $S_{n}$, is $R_{n+1}:R_n = 1:0.3$. 
After including the effects of attenuation, it is seen that the more distant the source, 
the earlier the GZK cutoff appears. This result may be easily understood through an application
of the analytic description for proton propagation provided in the first lecture.

The contribution from the different shells (also discussed in the previous lecture) to the photon flux and 
photon fraction are shown, respectively, in the left panel of Fig.\ref{photon_results}.
For this result, the Clark radio background radiation and a single $\gamma\gamma^{\rm bg}$ interaction 
(ie. strong extragalactic magnetic fields) were assumed. From this figure it is clear that the UHE-photon 
flux is suppressed for photons coming from regions farther than tens of Mpc away. 
The UHE-photon fraction for this case, shown in the right-panel of Fig.~\ref{photon_results}, 
shows two peaks with values $10^{-3}$ and $10^{-2}$, respectively, with the dominant contributing shell being 
9-27~Mpc.

To model a local over-density of UHECR accelerators, a ratio of sources from different shells 
$R_{n+1}:R_n=1:0.5$ and $R_{n+1}:R_n=1:0.7$ was adopted. Such an over-density leads to a less abrupt and slightly
higher in energy GZK suppression and an increase in the photon fraction at energies around the first peak.
Similarly, to describe an under-density in the UHECR accelerator distribution, a local void of UHECR 
sources was introduced. Such a local void lead to a slightly lower in energy and very abrupt
GZK suppression and a decrease in the photon fraction at energies around the first peak. For
this case an increase in the photon fraction at energies around the second peak was also observed.
However, such a photon component would not be detectable with present and (foreseeable) next generation
UHECR detectors.

The effect of diffusion on these results is to increase the path length of the protons relative
to that of the same energy photons, reducing the flux contribution from distant sources and thus, 
in a homogeneous scenario, increasing the photon fraction component. In this sense the photon fraction
values calculated can be considered conservative values.

\begin{figure}
\includegraphics[width=2.3in,angle=-90]{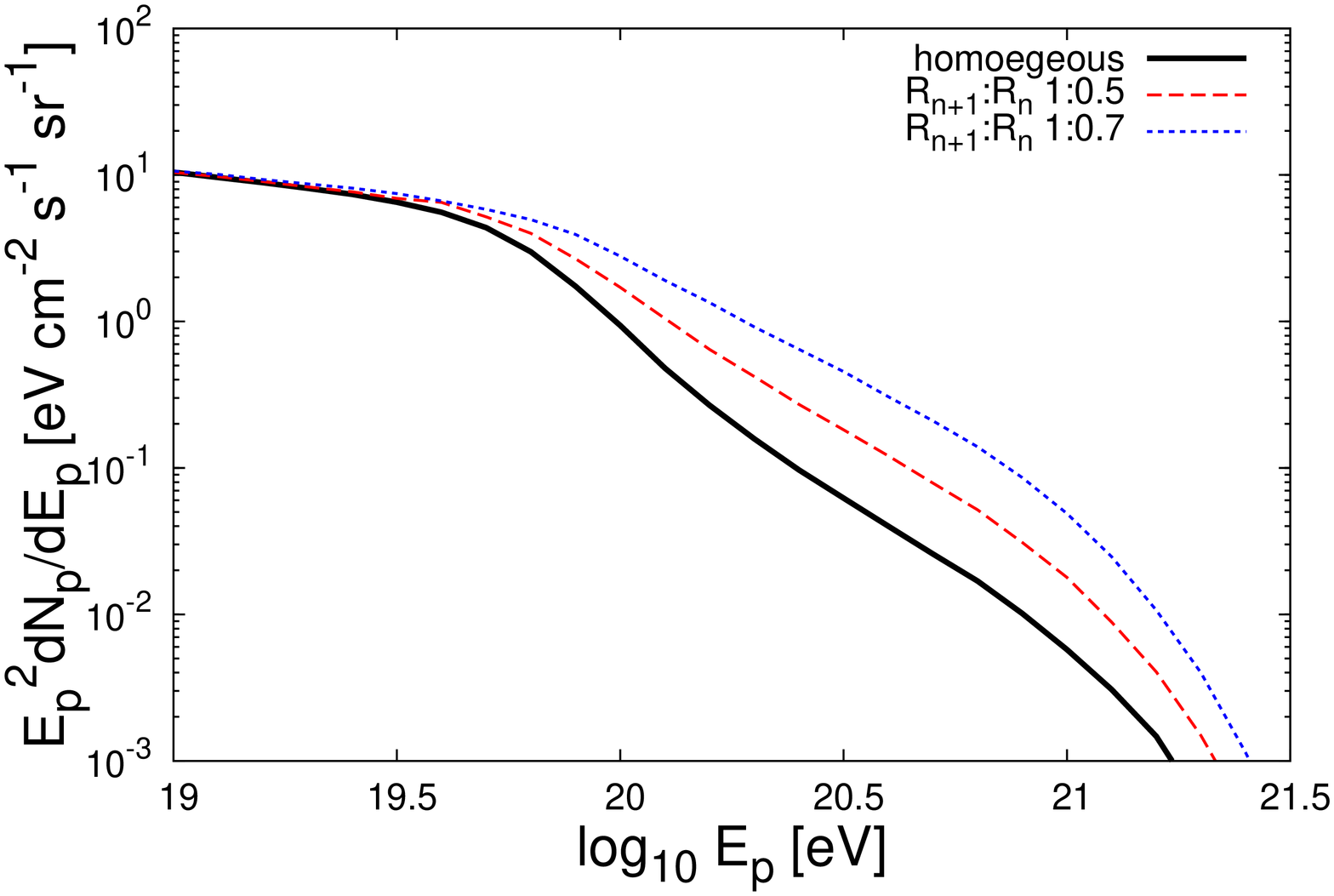}
\includegraphics[width=2.3in,angle=-90]{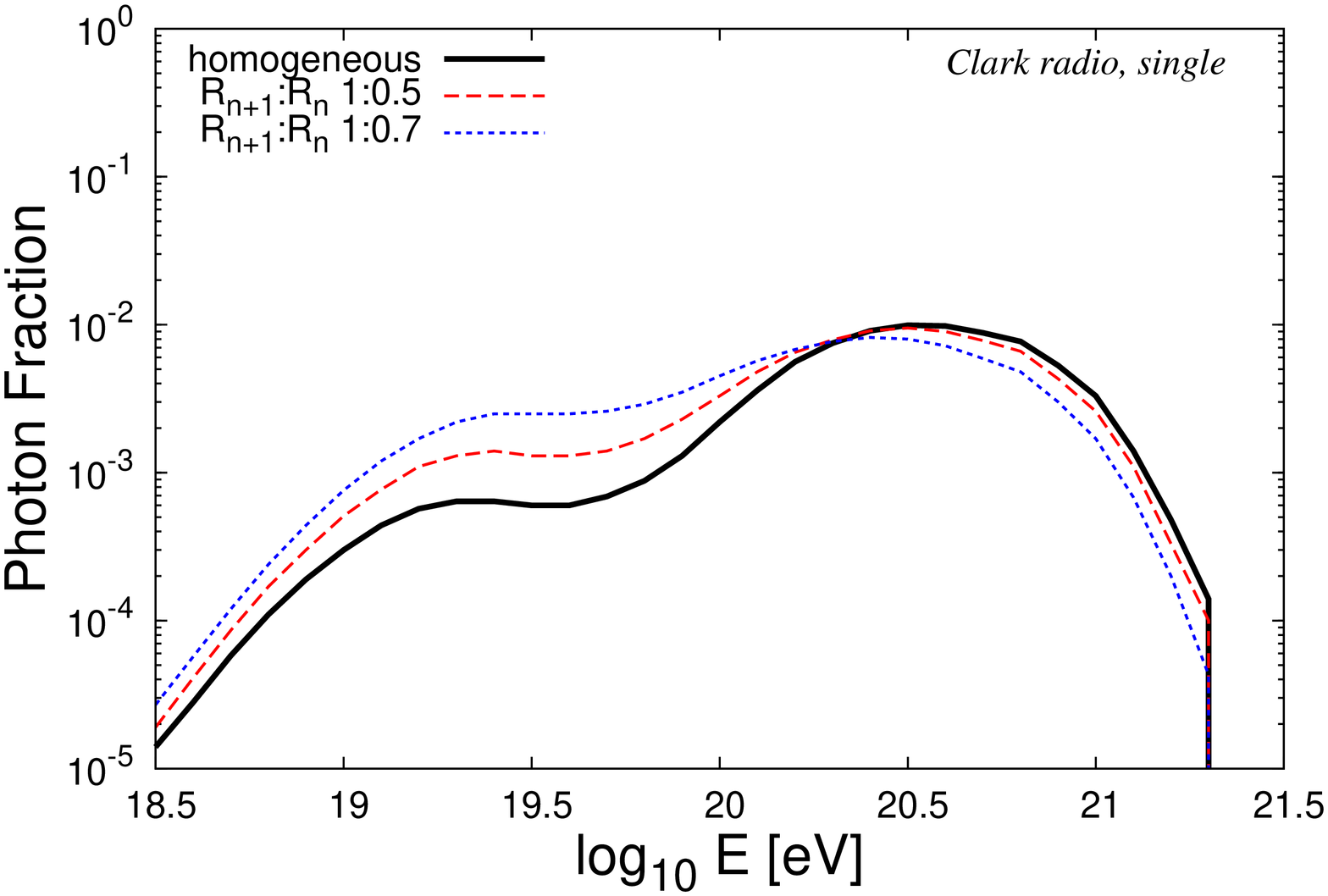}
\caption{{\bf Left}: The UHECR proton flux arriving for the case of a local over-density of UHECR sources distribution of sources for the cases of $R_{n+1}:R_{n}$ of 0.5 and 0.7. {\bf Right}: The corresponding photon fraction of the arriving UHECR obtained for the local over-density of UHECR sources for the cases of $R_{n+1}:R_{n}$ of 0.5 and 0.7. For this plot, the label ``{\it Clark radio, single}'' refers to the assumptions about the radio background and magnetic field strengths, for which we have assumed the Clark et al. \cite{Parker:1970} radio background and extra-galactic magnetic fields stronger than 0.1~nG. The injected proton spectrum assumed for these results had a spectral index of $\alpha$=2.}
\label{overdensity}
\end{figure}

\section{4.2 UHECR Nuclei and the Diffuse Neutrino Flux From Sources}
We here go through ``generic'' calculations for the possible neutrino flux from AGN, GRBs and starburst galaxies, 
making use of the source opacity, $f_{\pi}$, for $p\gamma$ interactions. 
At energies above $10^{15}$~eV, the sources of CR remain completely unknown. However, a list of candidates sources 
may be drawn up from the Hillas criteria \cite{Hillas}. Acceleration mechanisms aside, a simple dimensional 
argument of the possible acceleration sites and the maximum obtainable energy from these sites due to the magnetic 
field and the size of the site can be considered, giving
$$
\frac{E_{\rm max}}{10^{18}~{\rm eV}}=Ze\left(\frac{B}{\mu{\rm G}}c\right)\frac{R_{\rm source}}{\rm kpc},
$$
where $E_{\rm max}$ is the maximum energy obtainable by a particle in the source, $B$ is the magnetic field
in the source, and $R_{\rm source}$ is the size of the source region.
Together with a consideration of the physics to describe the escape time of the particle and the energy gain 
rate such as that achieved through diffusive shock acceleration models under the Bohm diffusion assumption,
for which the escape time $\sim R^{2}/(R_{\rm Larmor}c)$ and the acceleration time $\sim R/(\beta_{sh}^{2}c)$,
a refined description for the maximum energy is obtained,
$$
E_{\rm max}=\beta_{\rm sh}Ze(Bc)R_{\rm source},
$$
where $\beta_{\rm sh}$ is the velocity, in units of the speed of light in a vacuum ($c$), of the shock propagating through 
the source. Possible accelerators, able to achieve the particle energies observed at the high energy end of 
the spectrum ($\sim 10^{20}$~eV) are active Galactic nuclei (AGN) with $R\sim10^{-2}$~pc and $B\sim$~G, Gamma Ray Bursts 
(GRB) with $R\sim 10^{-6}$~pc and $B\sim 10^{3}$~G and starburst regions with $R\sim 100$~pc and $B\sim 100$~$\mu$G.

The luminosity break energy at the candidate sources is $10^{44}$~erg~s$^{-1}$ for AGN, 
$10^{52}$~erg~s$^{-1}$ for Gamma Ray Bursts (GRBs) and $10^{42}$~erg~s$^{-1}$ for starburst galaxies. 
These numbers must then be compared with the luminosity of a UHECR accelerator.
With UHECR above 10$^{18}$~eV having an energy density $\sim$10$^{-8}$~eV~cm$^{-3}$
or $10^{54}$~erg~Mpc$^{-3}$, the flux density of the sources, outputting for a
Hubble time must be 10$^{37}$~erg~Mpc$^{-3}$~s$^{-1}$. With source densities $\sim 10^{-5}$~Mpc$^{-3}$,
this puts a requirement of $L\sim 10^{42}$~erg~s$^{-1}$. Thus, all the sources considered
it seems may be considered reasonable, physically motivated, candidates.

\noindent {\bf AGN}: In this model, UHECRs are accelerated within relativistic blobs 
of plasma moving along the AGN jet. Assuming jet Lorentz factors $\sim\Gamma \sim 30$, 
and acceleration regions of size $l_{\rm source} \sim \Gamma c\Delta t \sim 10^{-2}$~pc, 
$\Delta t$ is the typical duration of a flare, the photon density within the source region 
with photon energies at which the spectrum ($E_{\gamma}dN_{\gamma}/dE_{\gamma}$) is maximum,  is
$n_{\gamma}^{\rm max}=10^{16}~$cm$^{-3}$. This leads to expected opacity factors of the source of 
$f_{\pi}=l_{\rm source}/l_{\rm int.}\approx 500$ \cite{Anchordoqui:2007tn}, where $l_{\rm source}$ is the size 
of the source region and $l_{\rm int.}\approx 1/(n_\gamma\sigma_{p\gamma})$ is the interaction
length of the particle.

\noindent {\bf GRB}: GRBs are brief flashes of high energy radiation that can outshine 
any other source in the sky. It is believed that the energetics associated with a GRB 
are caused by the dissipation of the kinetic energy of a relativistic expanding plasma 
wind called a ``fireball''. CRs may be accelerated in the fireball's
shocks where the Lorentz factor are $\Gamma \sim 300$. During the post-fireball phase, 
due to the expansion of the spherical shock in the shell rest frame and the relative motion 
between the shell rest frame and the observer's frame, the size of the original object is 
$l_{\rm source} = \Gamma^2 c \Delta t \sim 10^{-6}$~pc, where $\Delta t$ is the smallest observable timescale
in the prompt emission. The radiation field spectrum is modelled according to 
$E_{\gamma}dN_\gamma/dE_\gamma \propto E^{-\beta}_\gamma$, with $\beta=0,1$ below and above the break energy 
$E_{\gamma,br}=1$~MeV , respectively \cite{Band:1993eg}. The number density for photons is found to be 
$n_\gamma^{\rm max} \sim 10^{17}$~cm$^{-3}$. This leads to expected opacity factors of the source of 
$f_{\pi}=l_{\rm source}/l_{\rm int.}\approx 0.5$ \cite{Anchordoqui:2007tn}.

\noindent {\bf Starburst Galaxy}: These are galaxies undergoing an episode of large scale 
star formation. Within the starburst region of size $L \sim 100$~pc, OB and red supergiant 
stars emit UV radiation, whereas the dust surrounding the starburst region produces infrared 
emission\cite{Anchordoqui:2007tn}. These components are describable by a blackbody and 
a greybody respectively. In this sources, the photon number density is 
$n_\gamma \sim 10^5$~cm$^{-3}$, much less than in AGN or GRBs. This leads to expected opacity factors of the source of 
$f_{\pi}=l_{\rm source}/l_{\rm int.}\approx 10^{-4}$ \cite{Anchordoqui:2007tn}.

Protons interact with the source ambient radiation to photo-produce pions when the photon energy
in the proton's rest frame is above $E_{\gamma,th}\sim 145$~MeV. 
The decay of the secondary neutrons and charged pions produce a flux of neutrinos (as discussed in lecture 2). 

Both for AGN and GRB, charged pion decay is the dominant contribution to the cosmic neutrino flux, 
which at its peak value is around 75\% of the Waxman-Bahcall bound. For the Starburst model, 
due to the small values of $f_{\pi}$, the neutrino flux is 4 orders of magnitude below the Waxman-Bahcall limit.

If UHECR sources accelerate heavy nuclei instead of protons, the resulting high energy neutrino spectrum 
will be suppressed compared to an all-proton composition at acceleration. At the source, a nucleus interacts 
with the radiation fields photo-disintegrating into its constituent nucleons, which may then themselves produce neutrinos 
through photo-pion interactions. Since the density of the radiation fields at the source determines the 
intensity of the associated neutrino flux, if most of the nuclei are broken into their nucleons, the 
flux of neutrinos will be very similar to that predicted for accelerated protons. On the other hand, 
if the accelerated nuclei leave the source without interacting, the cosmic neutrino flux will be 
dramatically suppressed.

We can apply the proton interaction rate (as discussed in the first lecture) 

\begin{equation}\label{R_p}
  R_{p\gamma} \approx \sigma_{p\gamma} 
  \int_{(\xi_{p \gamma}-\Delta_{p\gamma})/2\Gamma_{p}}^{(\xi_{p\gamma}+\Delta_{p\gamma})/2\Gamma_{p}} 
  d\epsilon_{\gamma}\frac{dn_{\gamma}(\epsilon_{\gamma})}{d\epsilon_{\gamma}} ,
\end{equation}

\noindent to photo-disintegration reactions

\begin{equation}\label{R_Fe}
  R_{Fe\gamma} \approx \sigma_{Fe\gamma} 
  \int_{(\xi_{Fe\gamma}-\Delta_{Fe\gamma})/2\Gamma_{p}}^{(\xi_{Fe\gamma}+\Delta_{Fe\gamma})/2\Gamma_{p}} 
  d\epsilon_{\gamma}\frac{dn_\gamma(\epsilon_{\gamma})}{d\epsilon_{\gamma}}.
\end{equation}

\noindent where $dn_\gamma/d\epsilon_\gamma$ is the number density spectrum of ambient photons, 
$\sigma_{p\gamma}\approx 0.5$~mb, $\xi_{p\gamma}\approx 310$~MeV, $\Delta_{p\gamma}\approx 100$~MeV, 
$\sigma_{Fe\gamma}\approx 80$~mb, $\xi_{Fe\gamma}\approx 18$~MeV and $\Delta_{Fe\gamma}\approx 5$~MeV. 
This gives a ratio between photo pion and photo-disintegration rates of

\begin{equation}\label{R_pFe}
  R_{Fe\gamma}(\Gamma_{p})\approx  \frac{\sigma_{Fe\gamma}}{\sigma_{p\gamma}}
  R_{p\gamma}(15~\Gamma_{p})= 160 R_{p\gamma}(15~\Gamma_{p})\, .
\end{equation}
Complete photo-disintegration only occurs in GRBs for $E_{Fe}>10^{17}$~eV, and in AGN, 
for $E_{Fe}>10^{20}$~eV. The photo-disintegration at Starburst regions is negligible. 
As expected from this expression, in all cases the photo-disintegration curve 
followed the shape of the proton interaction rate, with the shape of both curves
being dictated by the shape of the radiation field spectrum.

\section {4.3 Conclusions}
In the first part of this lecture, the detection of the photon fraction component of UHECR spectrum 
has been demonstrated to be a powerful diagnostic tool for both verifying the origin of the suppression
feature observed of the in the UHECR spectrum as well as being usable, in conjunction with the UHECR 
cut-off feature observed, in determining the local distribution of UHECR sources.
In the second part of the lecture, generic calculation for the diffuse UHE-neutrino spectra from
candidate sources of UHECR (protons) have been gone through. Through the assumption, in both GRB and AGN
models, for near unity neutrino production opacities ($f_{\pi}$), complete disintegration of UHECR
nuclei in these objects has been demonstrated to be expected fpr energies above $10^{17}$~eV and
$10^{20}$~eV respectively. A consistency check of these opacities factors, whose values dictate the 
probability of UHECR protons interacting and producing UHE-neutrinos in the sources before escaping,
with the presence of Fe in the arriving UHECRs demands that much smaller neutrino fluxes from
these sources should be expected than obtained from present calculations which ignore the information
carried by the existence of UHECR nuclei.

\begin{theacknowledgments}
A.~T. acknowledges a research stipendium. E.~C. acknowledges the Organizing Comitee of the 3rd School on Cosmic Rays and Astrophysics for the finantial support. 
\end{theacknowledgments}

\end{document}